\documentclass[prb,lengthcheck,superscriptaddress]{revtex4-1}

\usepackage{amsmath}    
\usepackage{graphicx,epstopdf}    
\usepackage{verbatim}   
\usepackage{color}      
\usepackage{subfigure}  
\usepackage{hyperref}   
\usepackage{latexsym}
\usepackage[normalem]{ulem}

\newcommand{\fzd} {Helmholtz-Zentrum Dresden-Rossendorf, Institute of Ion Beam Physics and Materials Research, P.O. Box 510119, 01314 Dresden, Germany}

\newcommand{\als} {Advanced Light Source, Lawrence Berkeley National Laboratory, Berkeley, California 94720, USA}

\newcommand{\bessy} {Helmholtz-Zentrum Berlin f$\ddot{u}$r Materialien und Energie, Lise-Meitner-Campus, Hahn-Meitner-Platz 1, 14109 Berlin, Germany}
\newcommand{\tud} {Technische Universit\"{a}t Dresden, 01062 Dresden, Germany}
\newcommand{\unigrenoble}{Univ. Grenoble Alpes, INAC-SP2M, L$\_{Sim}$, F-38000 Grenoble, France} 
\newcommand{\ceagrenoble}{CEA, INAC-SP2M, Atomistic Simulation Lab., F-38000 Grenoble, France}
\newcommand{\TUC} {Faculty of Science, Technische Universit\"at Chemnitz, 09107 Chemnitz, Germany}
\newcommand{\cfaed} {Center for Advancing Electronics Dresden,  Technische Universit\"at Dresden, 01314 Dresden, Germany}

\begin{document}

\title{Revisiting defect-induced magnetism in graphite through neutron irradiation}

\author{Yutian Wang}
\address{\fzd}
\address{\tud}
\author{Pascal Pochet}
\address{\unigrenoble} 
\address{\ceagrenoble} 
\author{Catherine A. Jenkins}
\author{Elke Arenholz}
\address{\als}
\author{Gregor Bukalis}
\address{\bessy}
\author{Sibylle Gemming}
\address{\fzd}
\address{\TUC}
\address{\cfaed}
\author{Manfred Helm}
\address{\fzd}
\address{\tud}
\address{\cfaed}
\author{Shengqiang Zhou}
\address{\fzd}

\begin{abstract}
We have investigated the variation in the magnetization of highly ordered pyrolytic graphite (HOPG) after neutron irradiation, which introduces defects in the bulk sample and consequently gives rise to a large magnetic signal. We observe strong paramagnetism in HOPG, increasing with the neutron fluence. The induced paramagnetism can be well correlated with structural defects by comparison with density-functional theory calculations. In addition to the in-plane vacancies, the trans-planar defects also contribute to the magnetization. The lack of any magnetic order between the local moments is possibly due to the absence of hydrogen/nitrogen chemisorption, or the magnetic order cannot be established at all in the bulk form. 

\end{abstract}

\maketitle
\section{Introduction}
Defect induced magnetism in carbon based materials gives many
attractive perspectives in the fundamental understanding of
magnetism as well as in future spintronic applications. As early
as 2003 highly ordered pyrolytic graphite (HOPG) was reported to be ferromagnetic after proton
irradiation \cite{PhysRevLett.91.227201}, which provides an approach to control the defect-induced magnetism in graphite both concerning strength and in lateral distribution. After that, successive
investigations were performed for testing the reliability of the ferromagnetism in graphite
\cite{PhysRevLett.95.097201,lee2006electron,ISI:000262292200007,ISI:000271895500020,Yang20091399,PhysRevB.83.085417,He20111931,Shukla20121817} and for finding other carbon-based ferromagnetic materials
\cite{Makarova20031575,han2003observation,PhysRevB.72.224424,PhysRevB.75.075426,hohne2007magnetic,ISI:000302630100016}. As a consequence, the investigation on defect induced magnetism in semiconductors has been greatly stimulated  \cite{xing2009strong,PhysRevB.79.113201,PhysRevLett.99.127201,PhysRevLett.104.137201,PhysRevLett.106.087205,roever2011tracking}. 
So far experiments and theory show the following common
features:

\begin{enumerate}
    \item Paramagnetism can be greatly enhanced by introducing defects in graphite or graphene
\cite{PhysRevB.81.214404,ney2011irradiation,nair2012spin}. Some
research groups conclude that these paramagnetic centers do not
show any magnetic ordering down to 1.8 or 2 K
\cite{sepioni2010limits,ney2011irradiation,nair2012spin}.
    \item Ferromagnetism only appears under certain defect concentrations, i.e., in a narrow ion fluence window, and the
magnetization is weak
\cite{PhysRevB.79.113201,PhysRevB.81.214404,PhysRevB.85.144406,PhysRevLett.106.087205,Li201198}.
    \item In a microscopic picture, it has been found both theoretically \cite{PhysRevLett.99.107201} and experimentally \cite{PhysRevB.85.144406,PhysRevLett.104.096804} that defect-induced or disturbed electron states play an important
role in generating local moments in graphite.
    \item Foreign (or impurity) atoms, particularly, hydrogen and nitrogen, are helpful in establishing the ferromagnetic coupling between defects \cite{1367-2630-12-12-123012,PhysRevLett.99.107201,lehtinen2004irradiation}. 
\end{enumerate}

However, as to our knowledge, the research has focused mostly on thin-film like samples: ion implanted graphite with nm--$\mu$m affected thickness or graphene flakes. The as-measured magnetization is always in the range of 10$^{-6}$--10$^{-5}$ emu per sample \cite{PhysRevLett.91.227201,PhysRevB.79.113201,PhysRevB.81.214404,PhysRevB.85.144406,PhysRevLett.106.087205,roever2011tracking}.
The small magnetization renders data interpretation controversial as shown in a recent intensive discussion on the potential contamination in graphite
\cite{Esquinazi20101156,0295-5075-97-4-47001,0295-5075-98-5-57006,sepioni2012reply,spemann2013trace,venkatesan2013structural} as well as on artificial effects in magnetometry \cite{sawicki2011sensitive,0022-3727-44-21-215001}. Moreover, the implanted ions, especially those that differ chemically from the substrate, will stay in the matrix as
foreign atoms and an interface will naturally form between the
implanted region and the untouched substrate. Both the interface and the implanted ions will
make it difficult to unambiguously identify the defect type and hamper the interpretation of
the mechanism for the observed magnetization. To avoid these
problems we use neutron irradiation. Neutrons have a much stronger
penetrating capability than ions and will generate defects throughout
the whole sample. In this way, the foreign ion effect and the interface
effect can be excluded in the present study. Therefore, the application of neutron
irradiation could be a promising method to clarify the long standing question regarding the
origin of the defect induced magnetism in graphite in the following aspects. 

\begin{itemize}
    \item To verify whether the defect induced paramagnetism or ferromagnetism is a bulk effect or only a surface effect;
    \item To make a correlation between magnetism and defects based on the strong magnetic signal and results from various structural analysis techniques.
\end{itemize}

Accordingly, our work has been performed in the following way. HOPG specimens were subjected to neutron irradiation, whereby the irradiation fluence is varied to induce defects in graphite from slight damage to near amorphization. The magnetic and structural properties have been measured by various techniques. The results were complemented with a theoretical interpretation of the role of in-plane defects from literature and from new first-principles calculations of magnetic states of trans-planar divacancy configurations. 

The paper is organized as follows. In section II all experimental methods employed will be described. Then the results will be presented in three sub-sections. In section III.A, we present the large paramagnetism induced by irradiation and its dependence on the neutron fluence. In section III.B and C, the defect type and its concentration evolution will be discussed based on Raman and X-ray absorption spectroscopy, respectively. In section IV, we attempt to correlate the induced paramagnetic centers with in-plane vacancies and trans-planar defects by reviewing the literature data as well as by first-principles calculations. In the end of the discussion section, we also explain why the magnetic coupling between the induced moments is missing. The paper is finished with a short conclusion. 

\section{Experimental methods}

In the experiment, the used graphite samples were highly oriented pyrolytic graphite (HOPG) with a grade of ZYA, which are generally referred as graphite in this manuscript. Neutron irradiation was performed at the reactor BER II (Position DBVK) at Helmholtz-Zentrum Berlin \cite{lin2003calibration}. During irradiation the temperature of the samples was less than 50 $^{\circ}$C (see ref. \onlinecite{wendler2012damage}). Four samples were irradiated with the fluences of 6.24$\times$10$^{17}$, 1.25$\times$10$^{18}$,
6.24$\times$10$^{18}$, and 3.12$\times$10$^{19}$ cm$^{-2}$, which
are named as 3H, 6H, 30H and 150H according to the irradiation
time of 3 hours, 6 hours, 30 hours, 150 hours, respectively. The mechanism to produce crystal lattice defects by neutron irradiation is the elastic or inelastic scattering between neutrons and target nuclei. If the target nucleus gets enough energy after scattering, it will irreversibly displace the lattice atom from its original site, resulting in vacancies and interstitials. The minimum energy required to displace a carbon atom in graphite is around 25 eV \cite{kelly2000irradiation}. Therefore, we only consider the epithermal (0.5 eV -- 100 keV) and fast neutrons (100 keV -- 20 MeV) \cite{bode1996instrumental} in calculating the fluence. The elastic scattering dominates when the energy is below 5.5 MeV in carbon and the nuclear reaction (inelastic scattering) only becomes appreciable when the energy is above 9 MeV \cite{kelly2000irradiation}.

Magnetometry was performed using a SQUID-VSM (Quantum Design). The
magnetic properties were measured regarding their dependences on magnetic field and on temperature. The structure
change is characterized by Raman spectroscopy which is sensitive to defects in the aromatic ring, the edge state, the
hybridization type, the interstitial ions, and also to the stacking orders, etc
\cite{pimenta2007studying}. The $\mu$-Raman system is
equipped with a 532 nm wavelength laser and a liquid nitrogen
cooled CCD detector working in backscattering geometry.
X-ray absorption spectroscopy (XAS)
will further detect the bonding state change resulting from
neutron irradiation. The variations of the magnetization, the
Raman scattering and the X-ray absorption at the carbon K-edge
depending on the irradiation fluence allow us to clearly correlate the density of vacancies interstitials with the magnetism in the neutron irradiated graphite.

\begin{figure} \center
\includegraphics[scale=0.32]{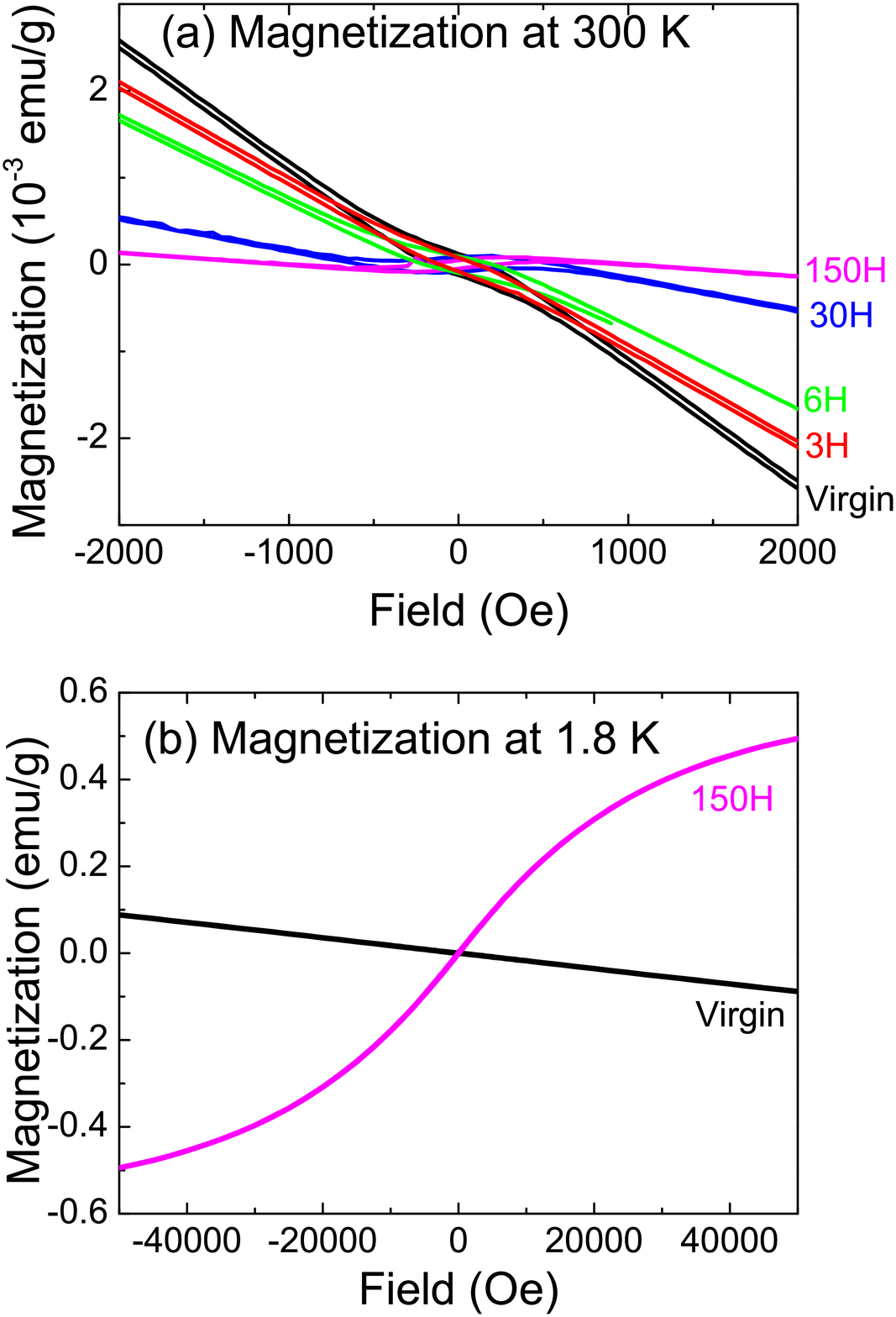}
\caption{Magnetization vs. field (a) the low field range at 300 K and (b) the large field range at 1.8 K.} \label{figFM_graphite}
\end{figure}

\section{Results and discussion}

\subsection{Magnetic properties}

\begin{figure} \center
\includegraphics[scale=0.7]{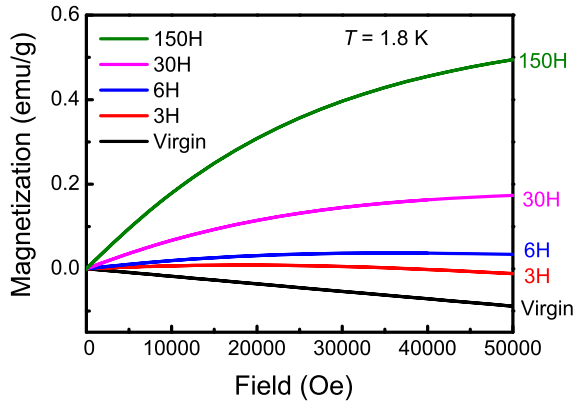}
\caption{The magnetic moments of all irradiated sample measured at
1.8 K as a function of the applied external field.} \label{FigMH_graphite}
\end{figure}

Figure \ref{figFM_graphite} shows the magnetization measurements at 300 K and 1.8 K for the virgin and irradiated graphite without any background correction. For the virgin graphite, the diamagnetic background
dominates the magnetic properties. A weak ferromagnetic hysteresis is observed already in the virgin
graphite. It is probably caused by intrinsic defects \cite{1367-2630-12-12-123012} or by Fe contamination
\cite{0295-5075-97-4-47001,0295-5075-98-5-57006,venkatesan2013structural}. Moreover, the ferromagnetic contribution is not changed significantly upon neutron irradiation. Therefore, this weak ferromagnetism is not the topic of our study in this manuscript. Besides the marginal change in the ferromagnetic component, there is a huge increment of the magnetization at low temperature. Figure \ref{figFM_graphite}(b) shows the comparison of the magnetization measurement at 1.8 K for the virgin graphite and sample 150H. Sample 150H shows a large paramagnetic component which will be discussed in detail later. Note that the change in the slope of the MH curves in Fig. \ref{figFM_graphite}(a) is due to the large increase of the paramagnetism upon irradiation as shown. At low temperature, the weak ferromagnetism in the irradiated samples is dominated by the paramagnetism and not resolvable. 

\begin{figure}[!htbp]

\centering
\includegraphics[scale=0.32]{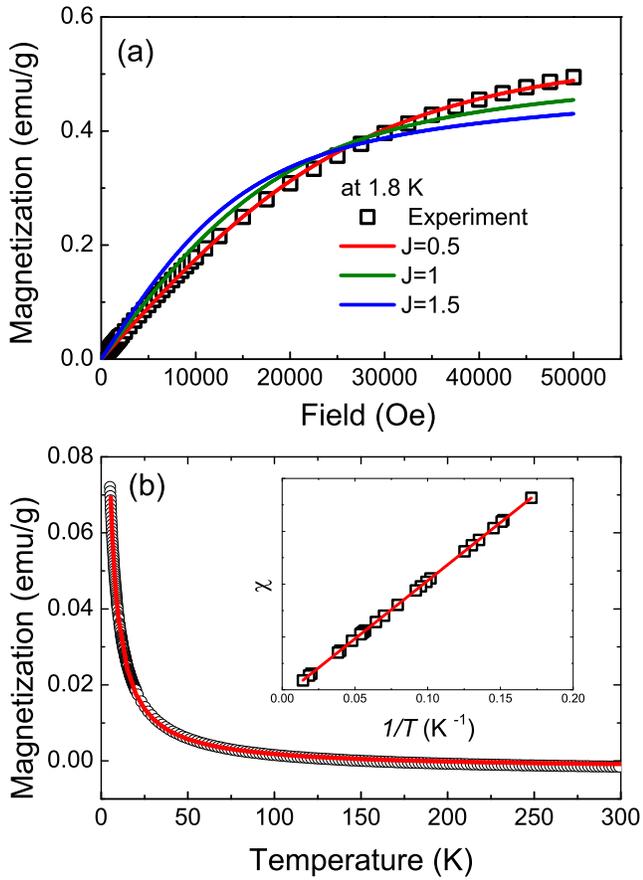}
\caption{(a) The measured magnetization at 1.8 K for sample 150H and the fitting using Brillouin function with \emph{J} = 0.5, 1, 1.5. (b) Temperature dependent susceptibility measured under a field of 10000 Oe. The black symbols are experimental data and the red solid curve is
is the fitting result by the equation (2). Inset: Inverse
susceptibility versus temperature demonstrating a linear, purely
paramagnetic behavior with no indication of magnetic ordering.}
\label{fig2fit_graphite}
\end{figure}

The field dependence of the magnetization at 1.8 K for all samples is shown in Fig.
\ref{FigMH_graphite}. Neutron irradiation leads to strong
paramagnetism. The graphite sample is changed completely
from diamagnetic-like to paramagnetic-like with increasing neutron
fluence. However, even for the sample with the highest neutron
fluence, the magnetization is not saturated at 1.8 K up to a field of 50000 Oe. In our experiment, the measured absolute magnetic moment for a graphite sample of around 4$\times$4 mm$^{2}$ is in the range of 0.001-0.01 emu at 1.8 or 5 K. This value is much larger than the previously reported ion implanted samples with a magnetic moment of around 10$^{-5}$--10$^{-6}$ emu \cite{PhysRevLett.91.227201,PhysRevB.79.113201,PhysRevB.81.214404,PhysRevB.85.144406} and is far above the sensitivity of SQUID-VSM. As shown in Fig. \ref{fig2fit_graphite}(a), the induced paramagnetism can
be precisely described by the standard Brillouin function after
removing the residual diamagnetic background and the intrinsic
paramagnetic contribution from the virgin graphite:

\begin{equation}\label{Brillouin}
M(\alpha)=NJ{\mu_B}g[\dfrac{2J+1}{2J}coth(\dfrac{2J+1}{2J}{\alpha})-\dfrac{1}{2J}coth(\dfrac{1}{2J}{\alpha})]
\end{equation}

where the $g$ factor is about 2 obtained from electron spin
resonance measurement (not shown), $\mu_B$ is Bohr magneton, $\alpha=gJ\mu_BH/k_BT$, $k_B$ is the the Boltzmann constant and \emph{N} is the density of spins. The Brillouin function provides excellent fits for J = 0.5, which corresponds to single electrons as charge carries and
$\textit{N}$ = 8$\times$10$^{19}$ $\mu_B$/mg for sample 150H. The fits using larger \emph{J}
unequivocally deviate from the shape of the measured M-H curves,
as they give significantly different, sharper changes with faster
saturation.

The Curie law
\begin{equation}\label{Curie}
    \chi=\frac{M}{H}=N\frac{J(J+1)(g\mu_B)^{2}}{3k_{B}T}
\end{equation}
with $J$ = 0.5 and $N$ = 8$\times$10$^{19}$ $\mu_B$/mg inferred
from Fig. \ref{fig2fit_graphite}(a) also gives a good fit to the temperature
dependent magnetization as shown in Fig. \ref{fig2fit_graphite}(b). The
inset of Fig. \ref{fig2fit_graphite}(b) shows the inverse susceptibility
versus temperature, revealing a linear, purely paramagnetic
behavior with no indication of magnetic ordering.

Figure \ref{Fig5MvsRaman} (shown later in the paper) shows the density of paramagnetic centers
obtained by fitting the magnetization measured at 1.8 K for
different samples as a function of neutron fluence in double
logarithmic scale. With increasing neutron fluence, i.e. the
amount of defects, more and more paramagnetic centers are
generated. This indicates that even the most strongly irradiated sample
is still not totally amorphous.

We also noted the work by Ramos et al. \cite{PhysRevB.81.214404}
Using ion implantation to introduce defects into graphite, they
reported an anomalous paramagnetic contribution. This
contribution remains independent of temperature up to 100 K, whereas the field dependent magnetization shows neither saturation nor any nonlinearity \cite{PhysRevB.81.214404}. Meanwhile
theoretical calculations also pointed out that if sufficient carbon
adatoms were available, they could weakly agglomerate in graphene and
superparamagnetism can be finally observed \cite{gerber2010first}.
However, in our experiment the magnetic properties for all samples
can be well described by spin 1/2 paramagnetism without superparamagnetic contributions. As expected if the whole volume contributes, in our experiment the as-measured magnetization signal
is as large as 0.001--0.01 emu per sample. The large magnetization
signal allows us to draw reliable conclusions and to exclude any
spurious and anomalous paramagnetic contribution.

\begin{figure}

\centering
\includegraphics[scale=0.35]{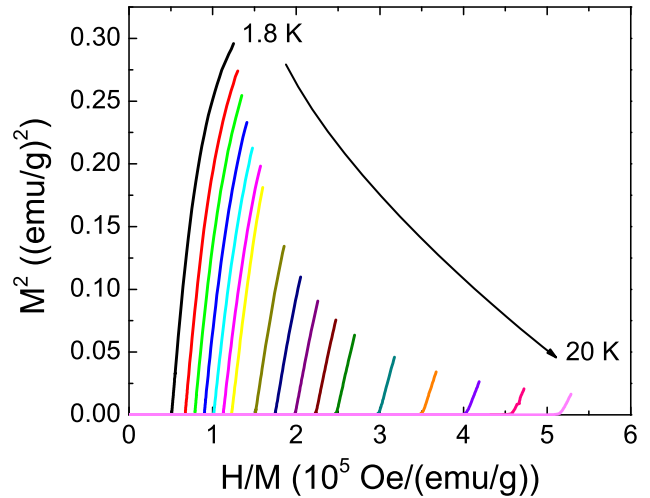}
\caption{Isothermal magnetization an Arrott plot: $M^2$ versus$H/M$. The lines with different color correspond to the measurements in the temperature range 1.8 to 20 K.}\label{FigArr}
\end{figure}

To further exclude a possible ferromagnetic ordering in our
sample we measured the magnetization vs. field at different
temperature to perform an Arrott plot analysis
\cite{arrott1957criterion}. This method is usually used to
accurately determine the Curie temperature $T_C$ and to verify the
paramagnetic to ferromagnetic phase transition. Such an analysis
is based on the relationship derived by Wohlfarth
\cite{wohlfarth1968very}
\begin{equation}\label{arr}
    [M(H,T)]^{2} = [M(0,0)]^{2}[1-(T/T_C)^{2}+2\chi_0H/M(H,T)]
\end{equation}
Note that this relationship results in parallel lines of the
isothermal $M^2$ which cross zero ($H/M=0$) in the vicinity of $T = T_C \pm \delta$.
Figure \ref{FigArr} shows the isothermal magnetization Arrott plot
for sample 150H (irradiated up to the highest fluence). The measurement
temperatures range from 1.8 K to 20 K. With increasing temperature,
the magnetization decreases, but none of the lines crosses the
zero point ($H/M=0$). It confirms that down to 1.8 K no magnetic order
appears in this sample. It is purely paramagnetic.


\begin{figure}
\includegraphics[scale=0.4]{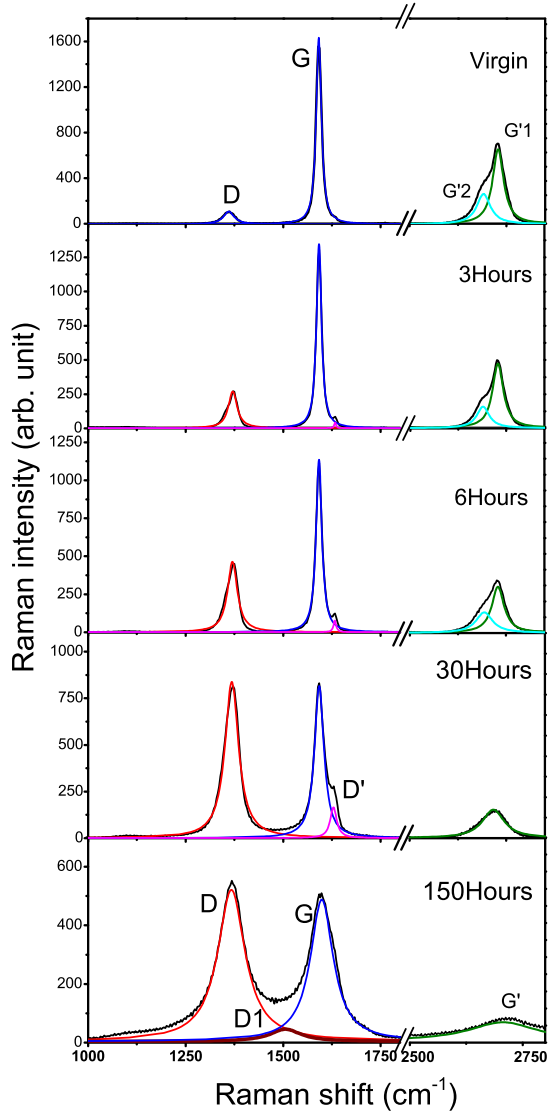}
\caption{Raman spectra of graphite after neutron irradiation. From the top to bottom shown are data for virgin graphite and
3 hours to 150 hours irradiated samples, respectively. The peaks
were deconvoluted to reveal the detailed variation after neutron
irradiation.} \label{Fig4Raman}
\end{figure}

\subsection{Raman spectroscopy}

Figure \ref{Fig4Raman} shows the Raman spectra of graphite samples
after neutron irradiation. From top to bottom are the virgin
sample and samples 3H...150H, respectively. A linear background
has been removed.

The reference sample shows the peaks typical for the high-quality HOPG \cite{pimenta2007studying,he2011raman}. The \emph{G} peak located at around 1590 cm$^{-1}$ corresponds to the inherent
\emph{E}$_{2g}$ mode of the aromatic ring. The \emph{D} peak
around 1360 cm$^{-1}$ represents an elastic scattering at defects in crystal
\cite{elman1982structural,pimenta2007studying,he2011raman,xing2013disorder}.

Upon neutron irradiation, the most pronounced changes occur in
the \emph{D} peak and in its overtone \emph{G'} peak (2D peak): the \emph{D} peak rises
with irradiation fluence and becomes as strong as the \emph{G} peak.
Two pronounced changes will be described in the following.

\subsubsection{In-plane vacancies}

\begin{figure}
\includegraphics[scale=0.32]{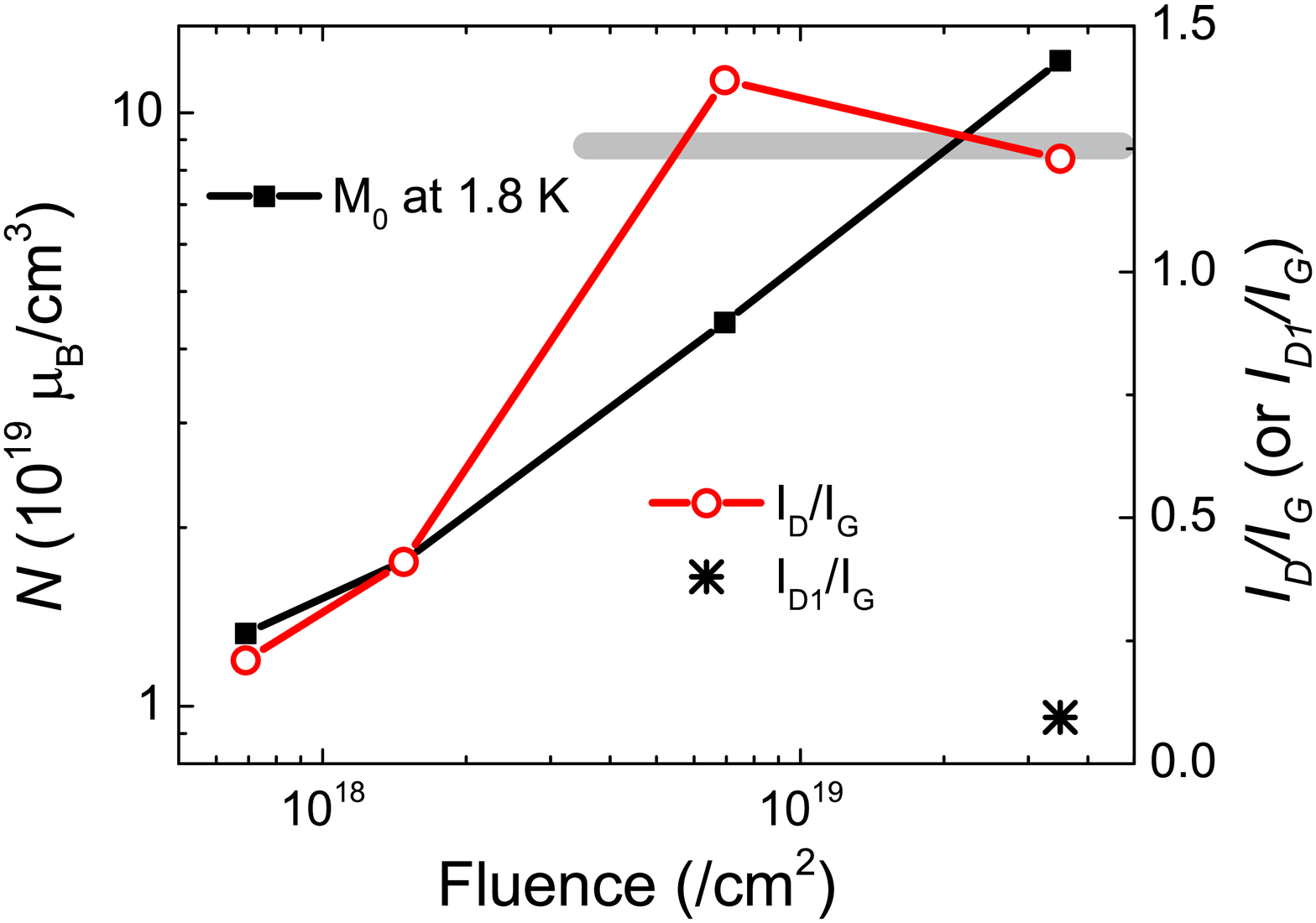}
\begin{flushleft}
\caption{The intensity ratio between \emph{D} and \emph{G} peaks (I$_D$/I$_G$) and
the fitted paramagnetic center density (\emph{N}, at 1.8 K) vs. neutron
irradiation fluence. The grey bar indicates the saturation value of I$_D$/I$_G$ for ion implanted graphite \cite{elman1982structural}.}\label{Fig5MvsRaman}
\end{flushleft}
\end{figure}

The increase of peak \emph{D} is generally attributed to the
in-plane vacancies in graphite
\cite{elman1982structural,pimenta2007studying,he2011raman,xing2013disorder}.
By independent methods such as X-ray diffraction and transmission
electron microscopy, the intensity ratio between \emph{D} and \emph{G} peaks
has been confirmed as a measure of the in-plane grain size. Neutron
irradiation induces a large number of interstitial and vacancy
pairs (\emph{I-V}). Most of \emph{I-V} defects will recombine
simultaneously and the remaining species can form various defects. Since a
high energy barrier blocks the diffusion of vacancies, most
vacancies become in-plane vacancies or form vacancy clusters. The
interstitial atoms prefer staying in the region between the layers
owing to the energetically highly unfavorable interstitial in-plane position
\cite{telling2007radiation}. In Figure \ref{Fig5MvsRaman}, we plot
the fluence dependent I$_{D}$/I$_{G}$ (the intensity ratio between \emph{D} and \emph{G} peaks). In our samples, the
strength of the \emph{D} peak increases with the neutron fluence
when the irradiation time is less than 30 hours. Further
increasing the neutron fluence, I$_{D}$/I$_{G}$ reaches a
saturation value. It indicates that with increasing the
irradiation time from 3 hours to 30 hours the density of vacancies
is continuously increasing until the vacancies reach a saturation
density. Such behavior was observed in ion irradiated or ball
milled graphite \cite{elman1982structural,xing2013disorder}. 

\subsubsection{Out-of-plane defects}

The \emph{G'} peak around 2720 cm$^{-1}$ is the overtone of the
\emph{D} peak. It is often referred as the 2D peak and is very
sensitive to the \emph{c}-axis stacking order of graphite. The
line shape and intensity of \emph{G'} are signatures of the
stacking of graphene layers. For bulk graphite consisting of an
...\emph{ABAB}...stacking, the \emph{G'} peak is composed of two
peaks. When the stacking is absent, the interaction between the
planes is very weak and they behave as two-dimensional crystals.
For a single graphene layer, the \emph{G'} peak is composed of a
single peak \cite{pimenta2007studying}. For our experiment, in the
virgin sample the interaction between the layers in 3\emph{D}
graphite makes the \emph{G'} peak to be split into
\emph{G'1} and \emph{G'2}. When the irradiation time is less than 6
hours, two peaks can fit the spectra, but their strength becomes
weak with increasing irradiation fluence. This indicates a
slight crystalline damage in the graphene sheet stacking. The influence of shear moments caused by
the interstitial atoms between the two sheets is less notable
for irradiation times of less than 6 hours. When the
irradiation time is over 30 hours, \emph{G'1} and \emph{G'2} peaks
decease strongly and mix into a single weak peak. This is
attributed to the out-of-plane defects in graphite
\cite{makarova2008ageing,he2011raman}. With increasing neutron
fluence, more interstitial atoms are assumed to diffuse into regions between the
graphene sheets so that the distance between the sheets increases strongly
enough, such that the graphene sheets behave like an isolated
single graphene sheet. The appearance of the \emph{D1} peak at
around 1500 cm$^{-1}$ for sample 150H is another indication for the
interstitial atoms between graphene sheets
\cite{jawhari1995raman,he2011raman}. At low fluence range, the
\emph{D1} peak is too weak to be fitted even for samples 30H. The
\emph{D1} peak was also observed in ion implanted graphite when the
implantation fluence is large enough \cite{he2011raman}.

This Raman analysis allows us to define two regimes for the four reported fluences. In the first regime (3H, 6H and 30H) defects are created in plane without interaction between neighboring planes. In the second regime (30H and 150H), the latter interaction becomes a dominant effect and trans-planar defects (interstitial or vacancy) are expected to play a major role: due to the high defect concentration, newly created defects are expected to combine with pre-existing defects in the neighboring planes as revealed by the rather saturated value of $I_D/I_G$ in the second fluence regime. Interestingly, these transplanar defects seem also contribute to the total magnetization.

\subsection{X-ray absorption spectroscopy}

To further probe the change in the electronic state in graphite
after neutron irradiation from a microscopic point of view, we
performed near-edge X-ray absorption fine structure spectroscopy
(NEXAFS, Beamline 6.3.1 at the Advanced Light Source in Berkeley).
The description of the experimental set up can be found in reference
\onlinecite{1367-2630-12-12-123012}. In our experiment, the
incident light was inclined by 45$^{\circ}$ to the sample surface. The signals
were collected in the total electron yield mode at room
temperature. All the spectra are normalized by the input flux for
comparison.

\begin{figure}
\includegraphics[scale=0.32]{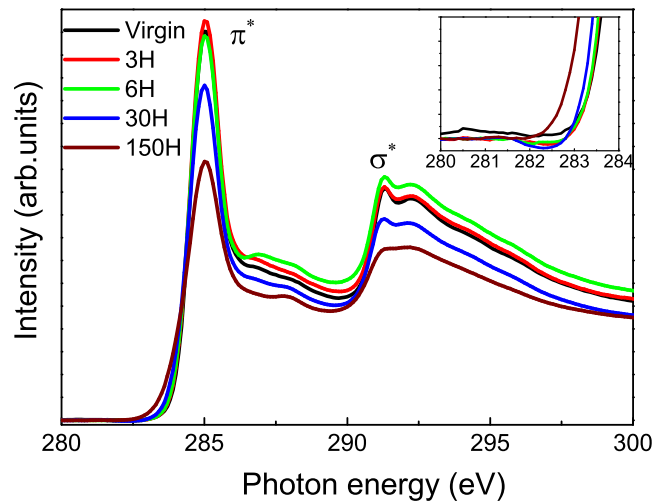}
\begin{flushleft}
\caption{The NEXAFS spectra of the graphite samples after neutron
irradiation for different time. Inset: zoom into the energy range
280--284 eV.}\label{Fig6XAS}
\end{flushleft}
\end{figure}

As shown in Fig. \ref{Fig6XAS}, there are two resonances around
285 eV and 292 eV, respectively. They correspond to the
transitions from \emph{1s} core-level electrons to $\pi^*$ and
$\sigma^*$ empty states, respectively. For samples 3H and 6H with
a small neutron fluence, there is no significant change either in
the peak intensity or in the peak shape compared with the virgin
sample. After the irradiation over 30 hours, the intensity of the $\pi^*$ peak decreases, 
which indicates that the aromatic $\pi$ system is severely perturbed. 
At the same time, the $\pi^*$ and $\sigma^*$ features are
becoming broader. In previous literature, it has been shown that
the $\pi^*$ and $\sigma^*$ resonances of carbon are much more
broadened in proton implanted graphite than our case 
\cite{PhysRevLett.98.187204,1367-2630-12-12-123012}.

The inset of Fig. \ref{Fig6XAS} shows a zoom into the energy range
280--284 eV. Compared with previous results on ion implanted graphite \cite{PhysRevB.85.144406}, the fundamental
difference of our sample is the missing of a pre-edge peak at
around 282 eV. In ref. \onlinecite{PhysRevB.85.144406},
a new small, but sizeable peak in the pre-edge region (281.5 eV
to 284.5 eV), has been reported in ion implanted ferromagnetic graphite.
This new peak was attributed to be closely related with defect states
near the Fermi energy level, and it was temporarily assigned to
rehybridized C-H bonds. The lack of rehybridized C-H bonds in our samples may explain the absence of ferromagnetism, which will be discussed later. 

\section{Discussion}

We have investigated the magnetic and structural properties of
graphite after neutron irradiation. Different from ion
implantation, neutron irradiation can introduce defects in the
whole graphite sample. The resulting magnetization is very large
and allows one to draw a reliable conclusion free of the influence of
contamination. Our experimental results lead to two conclusions: (1)
only spin 1/2 paramagnetism is induced in graphite by neutron
irradiation; and (2) both in-plane vacancies and out-of-plane
defects appear after irradiation. In this discussion, we attempt
to correlate the magnetization and defects and to understand why
the magnetic ordering is lacking.

\subsection{The origin of the paramagnetism}

Defect induced magnetism in both graphite and graphene has
been intensively investigated theoretically. Structural defects,
in general, can give rise to localized electronic states. It is
well accepted that the in-plane vacancies are the origin of 
local magnetic moments \cite{nair2012spin}. Upon removal of one atom, each of the
three neighboring atoms has one $sp^2$ dangling bond. Two of the C atoms
can form a pentagon, leaving one bond unsaturated. This remaining
dangling bond is responsible for the magnetic moment. Moreover,
the flat bands associated with defects lead to an increase in
the density of states at the Fermi level. Lehtinen et al., used
spin-polarized DFT and demonstrated that vacancies in graphite are
magnetic \cite{PhysRevLett.93.187202}. They also found that
hydrogen will strongly adsorb at vacancies in graphite,
maintaining the magnetic moment of the defect. Zhang et al.
\cite{PhysRevLett.99.107201} have confirmed that the local moments
appear near the vacancies and with increasing vacancy accumulation
the magnetization decreases non-monotonically. Using a combination
of a mean-field Hubbard model and first principles calculations,
Yazyev also confirmed that vacancies in graphite and graphene can
result in net magnetic moments \cite{yazyev2008magnetism}, while the
preserved stacking order of graphene layers is shown to be a
necessary condition for achieving a finite net magnetic moment of
irradiated graphite. In most calculations, the moment per
vacancy is sizeable up to 1--2
$\mu_B$\cite{PhysRevLett.93.187202,PhysRevLett.99.107201}. Indeed,
by scanning tunneling microscopy experiments, Ugeda \textit{et
al}. have observed a sharp electronic resonance at the Fermi
energy around a single vacancy in graphite, which can be
associated with the formation of local magnetic moments
\cite{PhysRevLett.104.096804}.

In our neutron irradiated graphite, we observed a strong
correlation between the magnetization and vacancies. Figure
\ref{Fig5MvsRaman} shows the irradiation-fluence dependent
magnetization and the values of $I_D$/$I_G$ of the Raman spectra.
At the low fluence regime, the density of magnetic moments shows
an excellent correlation with $I_D$/$I_G$ (the density of in-plane
vacancies): both increase monotonically with the fluence. This
indicates an agreement with the theoretical calculation: the
vacancy in graphite results in local magnetic moment. In the next
subsection, we discuss the role of out-of-plane defects.

\subsection{The role of trans-planar defects}

As shown in Fig. \ref{Fig5MvsRaman}, $I_D$/$I_G$ reaches its
saturation value of around 1.2--1.4 when the neutron fluence is higher.
$I_D$/$I_G$ of 1.2--1.4 is also a threshold of amorphisation in ion
irradiated graphite \cite{elman1982structural}. Despite the saturation in the density of
in-plane vacancies, the density of local moments still increases
with neutron fluence as shown in Fig.
\ref{Fig5MvsRaman}. What is the contribution for these additional local
magnetic moments? We consider the role of the trans-planar
defects. As shown in Fig. \ref{Fig4Raman}, for the largest
irradiation fluence, $D1$ peaks appears, which has been attributed to the trans-planar defects \cite{he2011raman}. In order to assess the experimental findings described in the above sections, we have investigated the possible magnetic state for trans-planar defects.

\begin{figure*}

\centering
\includegraphics[scale=0.8]{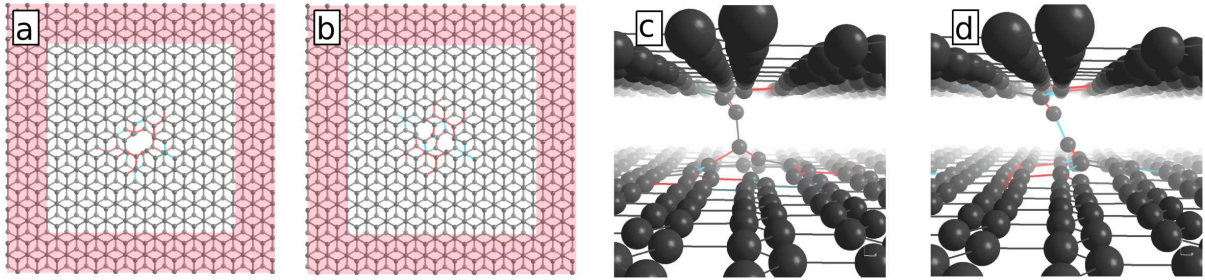}
\caption{ The top (left) and side (right) view of the two considered trans-planar divacancies: V$_{2}^{1}$ (a, c) and V$_{2}^{2}$ (b, d). Black balls are carbon atoms. Bonds between two neighboring atoms are colored as a function of length: black stands for standard distances [$2.70 \pm 0.05$ Bohr radius ($a_0$)], blue and red stand for short  ($2.60 \pm 0.05$ $a_0$) and  long  ($2.80 \pm 0.05$ $a_0$) distances, respectively.}\label{figDFT}
\end{figure*}

We start our analysis from the seminal work of Telling et al. \cite{Telling2003} who firstly propose the trans-planar divacancy configurations (see Figure \ref{figDFT}) that breaks the symmetry rules in graphite. Interestingly, the spin-polarized states for these defects was discussed in the paper but never assessed. In order to answer this question without any artifacts we have decided to run additional spin-polarized calculations in a super-cell which is large enough to avoid elastic effects between neighboring defect images (in-plane).
Systems containing 448 atoms per graphene sheets have proven to be reliable to study triangular vacancy clusters in hexagonal boron nitride sheets \cite{Machado2012} and are also used in the present study.
Here, two of these sheets with Bernal stacking were considered. The distance between the two sheets was fixed to 6.45 Bohr radii ($a_0$) for simplifying the treatment of the interlayer. This is achieved by the freezing of the perpendicular displacements in a band close to the edges of the super-cell (pink area in Fig. \ref{figDFT}). This treatment allows for a full relaxation both in-plane and out-of-plane of the central part of the super-cell where the defect sits. The PBE exchange and correlation function was chosen as it was found to well reproduce the in-plane relaxations  \cite{Krishnan2013}. The BigDFT \cite{Genovese2008} code was used to perform DFT calculations within surface boundary conditions \cite{Genovese2007}.

The two trans-planar divacancies V$_{2}^{2}$ and V$_{2}^{1}$ are considered together with the in-plane divacancy V$_{2}$ as a reference. The formation energy of the defect is calculated using the chemical potential of carbon in the pristine bilayer system. Singlet and triplet states are obtained by running spin averaged and spin polarized calculations, respectively. The results are summarized in table \ref{tabDFT}. The formation energy of the three defects increases in-line with the initial report of Telling et al. \cite{Telling2003}. However, important differences arise, underlying the role of the in-plane relaxations that were blocked in the previously used 64 atoms box \cite{Telling2003}. Indeed, while the estimated error of about 0.4 eV \cite{Telling2003} holds for the trans-planar vacancies, the difference is much more bigger for V$_{2}$. As a consequence, the energy difference between the two trans-planar divacancies remains in the order of 1.5 eV.

\begin{table}
\begin{center}
\caption{Formation energy (E$_f$) and energy difference (E$_{spin}$) between the singlet and triplet states for the three considered divacancies. The values in bracket are corresponding results in ref. \onlinecite{Telling2003}.
 \label{tabDFT}
}
\begin{tabular}{ccccc}
\hline
\hline
           Samples   & E$_f$ (eV) & E$_{spin}$ (meV)  \\
\hline
V$_2$ in-plane           & 7.55 (8.7)   & 2 \\
V$_{2}^{1}$ trans-planar & 13.85 (14.6) & 560\\
V$_{2}^{2}$ trans-planar & 12.77 (13.0) & 1 \\

\hline
\hline
\label{tabDFT}
\end{tabular}
\end{center}
\end{table}

In Table \ref{tabDFT} we also report the singlet to triplet formation energy for each defect.
In line with the report of a double bond \cite{Telling2003} for the inter-planar C-C bond (see bond length scale in figure \ref{figDFT}), the V$_{2}^{2}$ divacancy is in a singlet state.
This situation is different for the  V$_{2}^{1}$ divacancy: the inter-planar C-C bond is longer and more twisted, thus preventing further hybridization between the two carbon atoms.
As a consequence, the triplet state is stabilized by more than 500 meV with respect to the singlet state. According to Telling et al., the existence of triplet states gives a solid explanation for the observed spin 1/2 paramagnetism.


\subsection{Why is the magnetic interaction missing?}

As shown in Fig. \ref{FigMH_graphite} and Fig. \ref{Fig5MvsRaman}, the
paramagnetism in graphite can be strongly enhanced by irradiation
induced defects. After irradiation even to the largest fluence, the samples are not fully amorphous and $I_D$/$I_G$ of around 1.2 corresponds to a planar grain size of 3.5 nm \cite{elman1982structural}. Why is the magnetic interaction between the generated paramagnetic centers then missing? To answer
this question, we first need to estimate the density of defects,
i.e. the average distance between adjacent local moments.

Assuming the defects are homogeneously distributed in the sample
matrix, we estimate the average distances ($r$) between local
moments in our irradiated graphite samples. This value amounts to 2.2 nm for the sample with the largest neutron fluence. The nearest average distance between two spins is
around 16$a$ ($a$ = 0.14 nm is the C-C bond length).
Therefore, the direct coupling between the localized spins at the
vacancies is nearly negligible. Alternatively, the Ruderman-Kittel-Kasuya-Yosida
(RKKY) coupling is suggested to appear in defective graphite and
graphene \cite{PhysRevB75125408}. This coupling
might be ferromagnetic at a finite temperature when $k_Fr \ll 1$.
If assuming a Fermi energy of 20 meV in graphite
\cite{PhysRevB.76.161403}, the inverse of the Fermi wave vector
$1/k_F$ $\sim$ 30 nm. To have ferromagnetic ordering, the distance
between two spins $r$ should be $\ll$ 30 nm which corresponds to a spin density of 3.7$\times$10$^{18}$ cm$^{-3}$. In principle, all samples fulfill this criteria. All these moments may tend to be
ferromagnetically coupled via the RKKY coupling, although the Curie temperature can be very low
\cite{PhysRevB.76.161403}. However, we do not observe any magnetic ordering down to 1.8 K even for sample 150H.
It is not practical to further increase
the defect density, since the stacking order of graphenes plane must
be preserved \cite{yazyev2008magnetism,PhysRevB.85.144406}. 
Our sample with the highest neutron fluence is already at the verge
of amorphization. A larger irradiation fluence will perturb the
graphene lattice too much and destroy the necessary band structure
and carrier density.

Both published theory and experimental results suggest a crucial
role of hydrogen or nitrogen chemisorption in enhancing the spin
density and in establishing the magnetic coupling
\cite{PhysRevLett.93.187202,PhysRevLett.99.107201,yazyev2008magnetism,PhysRevB.76.161403,1367-2630-12-12-123012}.
All these moments from chemisorption will tend to be
ferromagnetically coupled, enhancing the Curie temperature by the
RKKY coupling. Recently, by careful angular dependent NEXAFS, He et al.
observed a new small peak in the pre-edge region (281.5 eV to
284.5 eV) \cite{PhysRevB.85.144406}. This new peak has been
interpreted to be closely related with the defect states near the
Fermi energy level and it is assigned to the formation of C-H bonds
\cite{1367-2630-12-12-123012}. Ohldag et al. also observed an X-ray
magnetic circular dichroism (XMCD) signal in the pre-edge region
of the C K-edge. However, as shown in Fig. \ref{Fig6XAS}, our present findings do not exhibit
any new peak in the pre-edge of the C K-edge. This may explain why the ferromagnetic coupling is missing.  

\section{Conclusion}

Neutron irradiation in graphite can induce a large amount of defects throughout the bulk specimens, consequently leading to a large measurable magnetization. This approach allows for a revisiting of defect induced magnetism in graphite by eliminating the influence of contamination or artificial effects. We conclude that only spin 1/2 paramagnetism is induced in neutron irradiated graphite. The creation of trans-planar vacancies (without dangling bonds) reduces the concentration of single in-plane vacancies. Complementing our study by first-principles calculations, we propose that both in-plane vacancies and trans-planar defects can form local magnetic moments, which are responsible for the observed 1/2 paramagnetism. The paramagnetism scales up with increasing the amount of defects, however, magnetic order unlikely can occur in a bulk form in defective graphite.

\section{Acknowledgement}

The work was financially supported by the Helmholtz-Gemeinschaft
Deutscher Forschungszentren (VH-NG-713 and VH-VI-442). Y. Wang thanks the
China Scholarship Council (File No. 2010675001) for supporting his
stay at HZDR. The authors also acknowledge the support by the International Science and Technology Cooperation Program of China (2012DFA51430). The Advanced Light Source is supported by the U.S. Department of Energy under Contract No.  DE-AC02-05CH11231. Calculations were performed using French HPC ressources from the GENCI-CCRT (grant 6194).


\begin{thebibliography}{63}%
\makeatletter
\providecommand \@ifxundefined [1]{%
 \@ifx{#1\undefined}
}%
\providecommand \@ifnum [1]{%
 \ifnum #1\expandafter \@firstoftwo
 \else \expandafter \@secondoftwo
 \fi
}%
\providecommand \@ifx [1]{%
 \ifx #1\expandafter \@firstoftwo
 \else \expandafter \@secondoftwo
 \fi
}%
\providecommand \natexlab [1]{#1}%
\providecommand \enquote  [1]{``#1''}%
\providecommand \bibnamefont  [1]{#1}%
\providecommand \bibfnamefont [1]{#1}%
\providecommand \citenamefont [1]{#1}%
\providecommand \href@noop [0]{\@secondoftwo}%
\providecommand \href [0]{\begingroup \@sanitize@url \@href}%
\providecommand \@href[1]{\@@startlink{#1}\@@href}%
\providecommand \@@href[1]{\endgroup#1\@@endlink}%
\providecommand \@sanitize@url [0]{\catcode `\\12\catcode `\$12\catcode
  `\&12\catcode `\#12\catcode `\^12\catcode `\_12\catcode `\%12\relax}%
\providecommand \@@startlink[1]{}%
\providecommand \@@endlink[0]{}%
\providecommand \url  [0]{\begingroup\@sanitize@url \@url }%
\providecommand \@url [1]{\endgroup\@href {#1}{\urlprefix }}%
\providecommand \urlprefix  [0]{URL }%
\providecommand \Eprint [0]{\href }%
\providecommand \doibase [0]{http://dx.doi.org/}%
\providecommand \selectlanguage [0]{\@gobble}%
\providecommand \bibinfo  [0]{\@secondoftwo}%
\providecommand \bibfield  [0]{\@secondoftwo}%
\providecommand \translation [1]{[#1]}%
\providecommand \BibitemOpen [0]{}%
\providecommand \bibitemStop [0]{}%
\providecommand \bibitemNoStop [0]{.\EOS\space}%
\providecommand \EOS [0]{\spacefactor3000\relax}%
\providecommand \BibitemShut  [1]{\csname bibitem#1\endcsname}%
\let\auto@bib@innerbib\@empty
\bibitem [{\citenamefont {Esquinazi}\ \emph {et~al.}(2003)\citenamefont
  {Esquinazi}, \citenamefont {Spemann}, \citenamefont {H\"ohne}, \citenamefont
  {Setzer}, \citenamefont {Han},\ and\ \citenamefont
  {Butz}}]{PhysRevLett.91.227201}%
  \BibitemOpen
  \bibfield  {author} {\bibinfo {author} {\bibfnamefont {P.}~\bibnamefont
  {Esquinazi}}, \bibinfo {author} {\bibfnamefont {D.}~\bibnamefont {Spemann}},
  \bibinfo {author} {\bibfnamefont {R.}~\bibnamefont {H\"ohne}}, \bibinfo
  {author} {\bibfnamefont {A.}~\bibnamefont {Setzer}}, \bibinfo {author}
  {\bibfnamefont {K.-H.}\ \bibnamefont {Han}}, \ and\ \bibinfo {author}
  {\bibfnamefont {T.}~\bibnamefont {Butz}},\ }\href {\doibase
  10.1103/PhysRevLett.91.227201} {\bibfield  {journal} {\bibinfo  {journal}
  {Phys. Rev. Lett.}\ }\textbf {\bibinfo {volume} {91}},\ \bibinfo {pages}
  {227201} (\bibinfo {year} {2003})}\BibitemShut {NoStop}%
\bibitem [{\citenamefont {Talapatra}\ \emph {et~al.}(2005)\citenamefont
  {Talapatra}, \citenamefont {Ganesan}, \citenamefont {Kim}, \citenamefont
  {Vajtai}, \citenamefont {Huang}, \citenamefont {Shima}, \citenamefont
  {Ramanath}, \citenamefont {Srivastava}, \citenamefont {Deevi},\ and\
  \citenamefont {Ajayan}}]{PhysRevLett.95.097201}%
  \BibitemOpen
  \bibfield  {author} {\bibinfo {author} {\bibfnamefont {S.}~\bibnamefont
  {Talapatra}}, \bibinfo {author} {\bibfnamefont {P.~G.}\ \bibnamefont
  {Ganesan}}, \bibinfo {author} {\bibfnamefont {T.}~\bibnamefont {Kim}},
  \bibinfo {author} {\bibfnamefont {R.}~\bibnamefont {Vajtai}}, \bibinfo
  {author} {\bibfnamefont {M.}~\bibnamefont {Huang}}, \bibinfo {author}
  {\bibfnamefont {M.}~\bibnamefont {Shima}}, \bibinfo {author} {\bibfnamefont
  {G.}~\bibnamefont {Ramanath}}, \bibinfo {author} {\bibfnamefont
  {D.}~\bibnamefont {Srivastava}}, \bibinfo {author} {\bibfnamefont {S.~C.}\
  \bibnamefont {Deevi}}, \ and\ \bibinfo {author} {\bibfnamefont {P.~M.}\
  \bibnamefont {Ajayan}},\ }\href {\doibase 10.1103/PhysRevLett.95.097201}
  {\bibfield  {journal} {\bibinfo  {journal} {Phys. Rev. Lett.}\ }\textbf
  {\bibinfo {volume} {95}},\ \bibinfo {pages} {097201} (\bibinfo {year}
  {2005})}\BibitemShut {NoStop}%
\bibitem [{\citenamefont {Lee}\ and\ \citenamefont
  {Lee}(2006)}]{lee2006electron}%
  \BibitemOpen
  \bibfield  {author} {\bibinfo {author} {\bibfnamefont {K.~W.}\ \bibnamefont
  {Lee}}\ and\ \bibinfo {author} {\bibfnamefont {C.~E.}\ \bibnamefont {Lee}},\
  }\href@noop {} {\bibfield  {journal} {\bibinfo  {journal} {Phys. Rev. Lett.}\
  }\textbf {\bibinfo {volume} {97}},\ \bibinfo {pages} {137206} (\bibinfo
  {year} {2006})}\BibitemShut {NoStop}%
\bibitem [{\citenamefont {Xia}\ \emph {et~al.}(2008)\citenamefont {Xia},
  \citenamefont {Li}, \citenamefont {Song}, \citenamefont {Yang}, \citenamefont
  {Liu}, \citenamefont {Zhao}, \citenamefont {Xia}, \citenamefont {Song},
  \citenamefont {Wang}, \citenamefont {Zhu}, \citenamefont {Gong},\ and\
  \citenamefont {Zhu}}]{ISI:000262292200007}%
  \BibitemOpen
  \bibfield  {author} {\bibinfo {author} {\bibfnamefont {H.}~\bibnamefont
  {Xia}}, \bibinfo {author} {\bibfnamefont {W.}~\bibnamefont {Li}}, \bibinfo
  {author} {\bibfnamefont {Y.}~\bibnamefont {Song}}, \bibinfo {author}
  {\bibfnamefont {X.}~\bibnamefont {Yang}}, \bibinfo {author} {\bibfnamefont
  {X.}~\bibnamefont {Liu}}, \bibinfo {author} {\bibfnamefont {M.}~\bibnamefont
  {Zhao}}, \bibinfo {author} {\bibfnamefont {Y.}~\bibnamefont {Xia}}, \bibinfo
  {author} {\bibfnamefont {C.}~\bibnamefont {Song}}, \bibinfo {author}
  {\bibfnamefont {T.-W.}\ \bibnamefont {Wang}}, \bibinfo {author}
  {\bibfnamefont {D.}~\bibnamefont {Zhu}}, \bibinfo {author} {\bibfnamefont
  {J.}~\bibnamefont {Gong}}, \ and\ \bibinfo {author} {\bibfnamefont
  {Z.}~\bibnamefont {Zhu}},\ }\href {\doibase 10.1002/adma.200801205}
  {\bibfield  {journal} {\bibinfo  {journal} {Adv. Mater.}\ }\textbf {\bibinfo
  {volume} {20}},\ \bibinfo {pages} {4679} (\bibinfo {year}
  {2008})}\BibitemShut {NoStop}%
\bibitem [{\citenamefont {\v{C}ervenka}\ \emph {et~al.}(2009)\citenamefont
  {\v{C}ervenka}, \citenamefont {Katsnelson},\ and\ \citenamefont
  {Flipse}}]{ISI:000271895500020}%
  \BibitemOpen
  \bibfield  {author} {\bibinfo {author} {\bibfnamefont {J.}~\bibnamefont
  {\v{C}ervenka}}, \bibinfo {author} {\bibfnamefont {M.~I.}\ \bibnamefont
  {Katsnelson}}, \ and\ \bibinfo {author} {\bibfnamefont {C.~F.~J.}\
  \bibnamefont {Flipse}},\ }\href {\doibase 10.1038/NPHYS1399} {\bibfield
  {journal} {\bibinfo  {journal} {Nat. Phys.}\ }\textbf {\bibinfo {volume}
  {5}},\ \bibinfo {pages} {840} (\bibinfo {year} {2009})}\BibitemShut {NoStop}%
\bibitem [{\citenamefont {Yang}\ \emph {et~al.}(2009)\citenamefont {Yang},
  \citenamefont {Xia}, \citenamefont {Qin}, \citenamefont {Li}, \citenamefont
  {Dai}, \citenamefont {Liu}, \citenamefont {Zhao}, \citenamefont {Xia},
  \citenamefont {Yan},\ and\ \citenamefont {Wang}}]{Yang20091399}%
  \BibitemOpen
  \bibfield  {author} {\bibinfo {author} {\bibfnamefont {X.}~\bibnamefont
  {Yang}}, \bibinfo {author} {\bibfnamefont {H.}~\bibnamefont {Xia}}, \bibinfo
  {author} {\bibfnamefont {X.}~\bibnamefont {Qin}}, \bibinfo {author}
  {\bibfnamefont {W.}~\bibnamefont {Li}}, \bibinfo {author} {\bibfnamefont
  {Y.}~\bibnamefont {Dai}}, \bibinfo {author} {\bibfnamefont {X.}~\bibnamefont
  {Liu}}, \bibinfo {author} {\bibfnamefont {M.}~\bibnamefont {Zhao}}, \bibinfo
  {author} {\bibfnamefont {Y.}~\bibnamefont {Xia}}, \bibinfo {author}
  {\bibfnamefont {S.}~\bibnamefont {Yan}}, \ and\ \bibinfo {author}
  {\bibfnamefont {B.}~\bibnamefont {Wang}},\ }\href {\doibase
  http://dx.doi.org/10.1016/j.carbon.2009.01.032} {\bibfield  {journal}
  {\bibinfo  {journal} {Carbon}\ }\textbf {\bibinfo {volume} {47}},\ \bibinfo
  {pages} {1399 } (\bibinfo {year} {2009})}\BibitemShut {NoStop}%
\bibitem [{\citenamefont {Makarova}\ \emph {et~al.}(2011)\citenamefont
  {Makarova}, \citenamefont {Shelankov}, \citenamefont {Serenkov},
  \citenamefont {Sakharov},\ and\ \citenamefont
  {Boukhvalov}}]{PhysRevB.83.085417}%
  \BibitemOpen
  \bibfield  {author} {\bibinfo {author} {\bibfnamefont {T.~L.}\ \bibnamefont
  {Makarova}}, \bibinfo {author} {\bibfnamefont {A.~L.}\ \bibnamefont
  {Shelankov}}, \bibinfo {author} {\bibfnamefont {I.~T.}\ \bibnamefont
  {Serenkov}}, \bibinfo {author} {\bibfnamefont {V.~I.}\ \bibnamefont
  {Sakharov}}, \ and\ \bibinfo {author} {\bibfnamefont {D.~W.}\ \bibnamefont
  {Boukhvalov}},\ }\href {\doibase 10.1103/PhysRevB.83.085417} {\bibfield
  {journal} {\bibinfo  {journal} {Phys. Rev. B}\ }\textbf {\bibinfo {volume}
  {83}},\ \bibinfo {pages} {085417} (\bibinfo {year} {2011})}\BibitemShut
  {NoStop}%
\bibitem [{\citenamefont {He}\ \emph {et~al.}(2011{\natexlab{a}})\citenamefont
  {He}, \citenamefont {Yang}, \citenamefont {Xia}, \citenamefont {Zhou},
  \citenamefont {Zhao}, \citenamefont {Song},\ and\ \citenamefont
  {Wang}}]{He20111931}%
  \BibitemOpen
  \bibfield  {author} {\bibinfo {author} {\bibfnamefont {Z.}~\bibnamefont
  {He}}, \bibinfo {author} {\bibfnamefont {X.}~\bibnamefont {Yang}}, \bibinfo
  {author} {\bibfnamefont {H.}~\bibnamefont {Xia}}, \bibinfo {author}
  {\bibfnamefont {X.}~\bibnamefont {Zhou}}, \bibinfo {author} {\bibfnamefont
  {M.}~\bibnamefont {Zhao}}, \bibinfo {author} {\bibfnamefont {Y.}~\bibnamefont
  {Song}}, \ and\ \bibinfo {author} {\bibfnamefont {T.}~\bibnamefont {Wang}},\
  }\href {\doibase http://dx.doi.org/10.1016/j.carbon.2011.01.018} {\bibfield
  {journal} {\bibinfo  {journal} {Carbon}\ }\textbf {\bibinfo {volume} {49}},\
  \bibinfo {pages} {1931 } (\bibinfo {year} {2011}{\natexlab{a}})}\BibitemShut
  {NoStop}%
\bibitem [{\citenamefont {Shukla}\ \emph {et~al.}(2012)\citenamefont {Shukla},
  \citenamefont {Sarkar}, \citenamefont {Banerji}, \citenamefont {Gupta},\ and\
  \citenamefont {Verma}}]{Shukla20121817}%
  \BibitemOpen
  \bibfield  {author} {\bibinfo {author} {\bibfnamefont {N.}~\bibnamefont
  {Shukla}}, \bibinfo {author} {\bibfnamefont {M.}~\bibnamefont {Sarkar}},
  \bibinfo {author} {\bibfnamefont {N.}~\bibnamefont {Banerji}}, \bibinfo
  {author} {\bibfnamefont {A.~K.}\ \bibnamefont {Gupta}}, \ and\ \bibinfo
  {author} {\bibfnamefont {H.~C.}\ \bibnamefont {Verma}},\ }\href {\doibase
  http://dx.doi.org/10.1016/j.carbon.2011.12.031} {\bibfield  {journal}
  {\bibinfo  {journal} {Carbon}\ }\textbf {\bibinfo {volume} {50}},\ \bibinfo
  {pages} {1817 } (\bibinfo {year} {2012})}\BibitemShut {NoStop}%
\bibitem [{\citenamefont {Makarova}\ \emph {et~al.}(2003)\citenamefont
  {Makarova}, \citenamefont {Han}, \citenamefont {Esquinazi}, \citenamefont
  {da~Silva}, \citenamefont {Kopelevich}, \citenamefont {Zakharova},\ and\
  \citenamefont {Sundqvist}}]{Makarova20031575}%
  \BibitemOpen
  \bibfield  {author} {\bibinfo {author} {\bibfnamefont {T.}~\bibnamefont
  {Makarova}}, \bibinfo {author} {\bibfnamefont {K.-H.}\ \bibnamefont {Han}},
  \bibinfo {author} {\bibfnamefont {P.}~\bibnamefont {Esquinazi}}, \bibinfo
  {author} {\bibfnamefont {R.}~\bibnamefont {da~Silva}}, \bibinfo {author}
  {\bibfnamefont {Y.}~\bibnamefont {Kopelevich}}, \bibinfo {author}
  {\bibfnamefont {I.}~\bibnamefont {Zakharova}}, \ and\ \bibinfo {author}
  {\bibfnamefont {B.}~\bibnamefont {Sundqvist}},\ }\href {\doibase
  http://dx.doi.org/10.1016/S0008-6223(03)00082-4} {\bibfield  {journal}
  {\bibinfo  {journal} {Carbon}\ }\textbf {\bibinfo {volume} {41}},\ \bibinfo
  {pages} {1575 } (\bibinfo {year} {2003})}\BibitemShut {NoStop}%
\bibitem [{\citenamefont {Han}\ \emph {et~al.}(2003)\citenamefont {Han},
  \citenamefont {Spemann}, \citenamefont {H{\"o}hne}, \citenamefont {Setzer},
  \citenamefont {Makarova}, \citenamefont {Esquinazi},\ and\ \citenamefont
  {Butz}}]{han2003observation}%
  \BibitemOpen
  \bibfield  {author} {\bibinfo {author} {\bibfnamefont {K.-H.}\ \bibnamefont
  {Han}}, \bibinfo {author} {\bibfnamefont {D.}~\bibnamefont {Spemann}},
  \bibinfo {author} {\bibfnamefont {R.}~\bibnamefont {H{\"o}hne}}, \bibinfo
  {author} {\bibfnamefont {A.}~\bibnamefont {Setzer}}, \bibinfo {author}
  {\bibfnamefont {T.}~\bibnamefont {Makarova}}, \bibinfo {author}
  {\bibfnamefont {P.}~\bibnamefont {Esquinazi}}, \ and\ \bibinfo {author}
  {\bibfnamefont {T.}~\bibnamefont {Butz}},\ }\href@noop {} {\bibfield
  {journal} {\bibinfo  {journal} {Carbon}\ }\textbf {\bibinfo {volume} {41}},\
  \bibinfo {pages} {785} (\bibinfo {year} {2003})}\BibitemShut {NoStop}%
\bibitem [{\citenamefont {Han}\ \emph {et~al.}(2005)\citenamefont {Han},
  \citenamefont {Talyzin}, \citenamefont {Dzwilewski}, \citenamefont
  {Makarova}, \citenamefont {H\"ohne}, \citenamefont {Esquinazi}, \citenamefont
  {Spemann},\ and\ \citenamefont {Dubrovinsky}}]{PhysRevB.72.224424}%
  \BibitemOpen
  \bibfield  {author} {\bibinfo {author} {\bibfnamefont {K.-H.}\ \bibnamefont
  {Han}}, \bibinfo {author} {\bibfnamefont {A.}~\bibnamefont {Talyzin}},
  \bibinfo {author} {\bibfnamefont {A.}~\bibnamefont {Dzwilewski}}, \bibinfo
  {author} {\bibfnamefont {T.~L.}\ \bibnamefont {Makarova}}, \bibinfo {author}
  {\bibfnamefont {R.}~\bibnamefont {H\"ohne}}, \bibinfo {author} {\bibfnamefont
  {P.}~\bibnamefont {Esquinazi}}, \bibinfo {author} {\bibfnamefont
  {D.}~\bibnamefont {Spemann}}, \ and\ \bibinfo {author} {\bibfnamefont
  {L.~S.}\ \bibnamefont {Dubrovinsky}},\ }\href {\doibase
  10.1103/PhysRevB.72.224424} {\bibfield  {journal} {\bibinfo  {journal} {Phys.
  Rev. B}\ }\textbf {\bibinfo {volume} {72}},\ \bibinfo {pages} {224424}
  (\bibinfo {year} {2005})}\BibitemShut {NoStop}%
\bibitem [{\citenamefont {Mathew}\ \emph {et~al.}(2007)\citenamefont {Mathew},
  \citenamefont {Satpati}, \citenamefont {Joseph}, \citenamefont {Dev},
  \citenamefont {Nirmala}, \citenamefont {Malik},\ and\ \citenamefont
  {Kesavamoorthy}}]{PhysRevB.75.075426}%
  \BibitemOpen
  \bibfield  {author} {\bibinfo {author} {\bibfnamefont {S.}~\bibnamefont
  {Mathew}}, \bibinfo {author} {\bibfnamefont {B.}~\bibnamefont {Satpati}},
  \bibinfo {author} {\bibfnamefont {B.}~\bibnamefont {Joseph}}, \bibinfo
  {author} {\bibfnamefont {B.~N.}\ \bibnamefont {Dev}}, \bibinfo {author}
  {\bibfnamefont {R.}~\bibnamefont {Nirmala}}, \bibinfo {author} {\bibfnamefont
  {S.~K.}\ \bibnamefont {Malik}}, \ and\ \bibinfo {author} {\bibfnamefont
  {R.}~\bibnamefont {Kesavamoorthy}},\ }\href {\doibase
  10.1103/PhysRevB.75.075426} {\bibfield  {journal} {\bibinfo  {journal} {Phys.
  Rev. B}\ }\textbf {\bibinfo {volume} {75}},\ \bibinfo {pages} {075426}
  (\bibinfo {year} {2007})}\BibitemShut {NoStop}%
\bibitem [{\citenamefont {H{\"o}hne}\ \emph {et~al.}(2007)\citenamefont
  {H{\"o}hne}, \citenamefont {Esquinazi}, \citenamefont {Heera},\ and\
  \citenamefont {Weishart}}]{hohne2007magnetic}%
  \BibitemOpen
  \bibfield  {author} {\bibinfo {author} {\bibfnamefont {R.}~\bibnamefont
  {H{\"o}hne}}, \bibinfo {author} {\bibfnamefont {P.}~\bibnamefont
  {Esquinazi}}, \bibinfo {author} {\bibfnamefont {V.}~\bibnamefont {Heera}}, \
  and\ \bibinfo {author} {\bibfnamefont {H.}~\bibnamefont {Weishart}},\
  }\href@noop {} {\bibfield  {journal} {\bibinfo  {journal} {Diam. Relat.
  Mater.}\ }\textbf {\bibinfo {volume} {16}},\ \bibinfo {pages} {1589}
  (\bibinfo {year} {2007})}\BibitemShut {NoStop}%
\bibitem [{\citenamefont {Ma}\ \emph {et~al.}(2012)\citenamefont {Ma},
  \citenamefont {Lu}, \citenamefont {Yi}, \citenamefont {Feng}, \citenamefont
  {Herng}, \citenamefont {Liu}, \citenamefont {Gao}, \citenamefont {Xue},
  \citenamefont {Xue}, \citenamefont {Ouyang},\ and\ \citenamefont
  {Ding}}]{ISI:000302630100016}%
  \BibitemOpen
  \bibfield  {author} {\bibinfo {author} {\bibfnamefont {Y.~W.}\ \bibnamefont
  {Ma}}, \bibinfo {author} {\bibfnamefont {Y.~H.}\ \bibnamefont {Lu}}, \bibinfo
  {author} {\bibfnamefont {J.~B.}\ \bibnamefont {Yi}}, \bibinfo {author}
  {\bibfnamefont {Y.~P.}\ \bibnamefont {Feng}}, \bibinfo {author}
  {\bibfnamefont {T.~S.}\ \bibnamefont {Herng}}, \bibinfo {author}
  {\bibfnamefont {X.}~\bibnamefont {Liu}}, \bibinfo {author} {\bibfnamefont
  {D.~Q.}\ \bibnamefont {Gao}}, \bibinfo {author} {\bibfnamefont {D.~S.}\
  \bibnamefont {Xue}}, \bibinfo {author} {\bibfnamefont {J.~M.}\ \bibnamefont
  {Xue}}, \bibinfo {author} {\bibfnamefont {J.~Y.}\ \bibnamefont {Ouyang}}, \
  and\ \bibinfo {author} {\bibfnamefont {J.}~\bibnamefont {Ding}},\ }\href
  {\doibase 10.1038/ncomms1689} {\bibfield  {journal} {\bibinfo  {journal}
  {Nature Commun.}\ }\textbf {\bibinfo {volume} {3}},\ \bibinfo {pages} {727}
  (\bibinfo {year} {2012})}\BibitemShut {NoStop}%
\bibitem [{\citenamefont {Xing}\ \emph {et~al.}(2009)\citenamefont {Xing},
  \citenamefont {Yi}, \citenamefont {Wang}, \citenamefont {Liao}, \citenamefont
  {Yu}, \citenamefont {Shen}, \citenamefont {Huan}, \citenamefont {Sum},
  \citenamefont {Ding},\ and\ \citenamefont {Wu}}]{xing2009strong}%
  \BibitemOpen
  \bibfield  {author} {\bibinfo {author} {\bibfnamefont {G.}~\bibnamefont
  {Xing}}, \bibinfo {author} {\bibfnamefont {J.}~\bibnamefont {Yi}}, \bibinfo
  {author} {\bibfnamefont {D.}~\bibnamefont {Wang}}, \bibinfo {author}
  {\bibfnamefont {L.}~\bibnamefont {Liao}}, \bibinfo {author} {\bibfnamefont
  {T.}~\bibnamefont {Yu}}, \bibinfo {author} {\bibfnamefont {Z.}~\bibnamefont
  {Shen}}, \bibinfo {author} {\bibfnamefont {C.}~\bibnamefont {Huan}}, \bibinfo
  {author} {\bibfnamefont {T.}~\bibnamefont {Sum}}, \bibinfo {author}
  {\bibfnamefont {J.}~\bibnamefont {Ding}}, \ and\ \bibinfo {author}
  {\bibfnamefont {T.}~\bibnamefont {Wu}},\ }\href@noop {} {\bibfield  {journal}
  {\bibinfo  {journal} {Phys. Rev. B}\ }\textbf {\bibinfo {volume} {79}},\
  \bibinfo {pages} {174406} (\bibinfo {year} {2009})}\BibitemShut {NoStop}%
\bibitem [{\citenamefont {Zhou}\ \emph {et~al.}(2009)\citenamefont {Zhou},
  \citenamefont {\ifmmode \check{C}\else \v{C}\fi{}i\ifmmode~\check{z}\else
  \v{z}\fi{}m\'ar}, \citenamefont {Potzger}, \citenamefont {Krause},
  \citenamefont {Talut}, \citenamefont {Helm}, \citenamefont {Fassbender},
  \citenamefont {Zvyagin}, \citenamefont {Wosnitza},\ and\ \citenamefont
  {Schmidt}}]{PhysRevB.79.113201}%
  \BibitemOpen
  \bibfield  {author} {\bibinfo {author} {\bibfnamefont {S.}~\bibnamefont
  {Zhou}}, \bibinfo {author} {\bibfnamefont {E.}~\bibnamefont {\ifmmode
  \check{C}\else \v{C}\fi{}i\ifmmode~\check{z}\else \v{z}\fi{}m\'ar}}, \bibinfo
  {author} {\bibfnamefont {K.}~\bibnamefont {Potzger}}, \bibinfo {author}
  {\bibfnamefont {M.}~\bibnamefont {Krause}}, \bibinfo {author} {\bibfnamefont
  {G.}~\bibnamefont {Talut}}, \bibinfo {author} {\bibfnamefont
  {M.}~\bibnamefont {Helm}}, \bibinfo {author} {\bibfnamefont {J.}~\bibnamefont
  {Fassbender}}, \bibinfo {author} {\bibfnamefont {S.~A.}\ \bibnamefont
  {Zvyagin}}, \bibinfo {author} {\bibfnamefont {J.}~\bibnamefont {Wosnitza}}, \
  and\ \bibinfo {author} {\bibfnamefont {H.}~\bibnamefont {Schmidt}},\ }\href
  {\doibase 10.1103/PhysRevB.79.113201} {\bibfield  {journal} {\bibinfo
  {journal} {Phys. Rev. B}\ }\textbf {\bibinfo {volume} {79}},\ \bibinfo
  {pages} {113201} (\bibinfo {year} {2009})}\BibitemShut {NoStop}%
\bibitem [{\citenamefont {Pan}\ \emph {et~al.}(2007)\citenamefont {Pan},
  \citenamefont {Yi}, \citenamefont {Shen}, \citenamefont {Wu}, \citenamefont
  {Yang}, \citenamefont {Lin}, \citenamefont {Feng}, \citenamefont {Ding},
  \citenamefont {Van},\ and\ \citenamefont {Yin}}]{PhysRevLett.99.127201}%
  \BibitemOpen
  \bibfield  {author} {\bibinfo {author} {\bibfnamefont {H.}~\bibnamefont
  {Pan}}, \bibinfo {author} {\bibfnamefont {J.~B.}\ \bibnamefont {Yi}},
  \bibinfo {author} {\bibfnamefont {L.}~\bibnamefont {Shen}}, \bibinfo {author}
  {\bibfnamefont {R.~Q.}\ \bibnamefont {Wu}}, \bibinfo {author} {\bibfnamefont
  {J.~H.}\ \bibnamefont {Yang}}, \bibinfo {author} {\bibfnamefont {J.~Y.}\
  \bibnamefont {Lin}}, \bibinfo {author} {\bibfnamefont {Y.~P.}\ \bibnamefont
  {Feng}}, \bibinfo {author} {\bibfnamefont {J.}~\bibnamefont {Ding}}, \bibinfo
  {author} {\bibfnamefont {L.~H.}\ \bibnamefont {Van}}, \ and\ \bibinfo
  {author} {\bibfnamefont {J.~H.}\ \bibnamefont {Yin}},\ }\href {\doibase
  10.1103/PhysRevLett.99.127201} {\bibfield  {journal} {\bibinfo  {journal}
  {Phys. Rev. Lett.}\ }\textbf {\bibinfo {volume} {99}},\ \bibinfo {pages}
  {127201} (\bibinfo {year} {2007})}\BibitemShut {NoStop}%
\bibitem [{\citenamefont {Yi}\ \emph {et~al.}(2010)\citenamefont {Yi},
  \citenamefont {Lim}, \citenamefont {Xing}, \citenamefont {Fan}, \citenamefont
  {Van}, \citenamefont {Huang}, \citenamefont {Yang}, \citenamefont {Huang},
  \citenamefont {Qin}, \citenamefont {Wang}, \citenamefont {Wu}, \citenamefont
  {Wang}, \citenamefont {Zhang}, \citenamefont {Gao}, \citenamefont {Liu},
  \citenamefont {Wee}, \citenamefont {Feng},\ and\ \citenamefont
  {Ding}}]{PhysRevLett.104.137201}%
  \BibitemOpen
  \bibfield  {author} {\bibinfo {author} {\bibfnamefont {J.~B.}\ \bibnamefont
  {Yi}}, \bibinfo {author} {\bibfnamefont {C.~C.}\ \bibnamefont {Lim}},
  \bibinfo {author} {\bibfnamefont {G.~Z.}\ \bibnamefont {Xing}}, \bibinfo
  {author} {\bibfnamefont {H.~M.}\ \bibnamefont {Fan}}, \bibinfo {author}
  {\bibfnamefont {L.~H.}\ \bibnamefont {Van}}, \bibinfo {author} {\bibfnamefont
  {S.~L.}\ \bibnamefont {Huang}}, \bibinfo {author} {\bibfnamefont {K.~S.}\
  \bibnamefont {Yang}}, \bibinfo {author} {\bibfnamefont {X.~L.}\ \bibnamefont
  {Huang}}, \bibinfo {author} {\bibfnamefont {X.~B.}\ \bibnamefont {Qin}},
  \bibinfo {author} {\bibfnamefont {B.~Y.}\ \bibnamefont {Wang}}, \bibinfo
  {author} {\bibfnamefont {T.}~\bibnamefont {Wu}}, \bibinfo {author}
  {\bibfnamefont {L.}~\bibnamefont {Wang}}, \bibinfo {author} {\bibfnamefont
  {H.~T.}\ \bibnamefont {Zhang}}, \bibinfo {author} {\bibfnamefont {X.~Y.}\
  \bibnamefont {Gao}}, \bibinfo {author} {\bibfnamefont {T.}~\bibnamefont
  {Liu}}, \bibinfo {author} {\bibfnamefont {A.~T.~S.}\ \bibnamefont {Wee}},
  \bibinfo {author} {\bibfnamefont {Y.~P.}\ \bibnamefont {Feng}}, \ and\
  \bibinfo {author} {\bibfnamefont {J.}~\bibnamefont {Ding}},\ }\href {\doibase
  10.1103/PhysRevLett.104.137201} {\bibfield  {journal} {\bibinfo  {journal}
  {Phys. Rev. Lett.}\ }\textbf {\bibinfo {volume} {104}},\ \bibinfo {pages}
  {137201} (\bibinfo {year} {2010})}\BibitemShut {NoStop}%
\bibitem [{\citenamefont {Liu}\ \emph {et~al.}(2011)\citenamefont {Liu},
  \citenamefont {Wang}, \citenamefont {Wang}, \citenamefont {Yang},
  \citenamefont {Chen}, \citenamefont {Qin}, \citenamefont {Song},
  \citenamefont {Wang},\ and\ \citenamefont {Chen}}]{PhysRevLett.106.087205}%
  \BibitemOpen
  \bibfield  {author} {\bibinfo {author} {\bibfnamefont {Y.}~\bibnamefont
  {Liu}}, \bibinfo {author} {\bibfnamefont {G.}~\bibnamefont {Wang}}, \bibinfo
  {author} {\bibfnamefont {S.}~\bibnamefont {Wang}}, \bibinfo {author}
  {\bibfnamefont {J.}~\bibnamefont {Yang}}, \bibinfo {author} {\bibfnamefont
  {L.}~\bibnamefont {Chen}}, \bibinfo {author} {\bibfnamefont {X.}~\bibnamefont
  {Qin}}, \bibinfo {author} {\bibfnamefont {B.}~\bibnamefont {Song}}, \bibinfo
  {author} {\bibfnamefont {B.}~\bibnamefont {Wang}}, \ and\ \bibinfo {author}
  {\bibfnamefont {X.}~\bibnamefont {Chen}},\ }\href {\doibase
  10.1103/PhysRevLett.106.087205} {\bibfield  {journal} {\bibinfo  {journal}
  {Phys. Rev. Lett.}\ }\textbf {\bibinfo {volume} {106}},\ \bibinfo {pages}
  {087205} (\bibinfo {year} {2011})}\BibitemShut {NoStop}%
\bibitem [{\citenamefont {Roever}\ \emph {et~al.}(2011)\citenamefont {Roever},
  \citenamefont {Malindretos}, \citenamefont {Bedoya-Pinto}, \citenamefont
  {Rizzi}, \citenamefont {Rauch},\ and\ \citenamefont
  {Tuomisto}}]{roever2011tracking}%
  \BibitemOpen
  \bibfield  {author} {\bibinfo {author} {\bibfnamefont {M.}~\bibnamefont
  {Roever}}, \bibinfo {author} {\bibfnamefont {J.}~\bibnamefont {Malindretos}},
  \bibinfo {author} {\bibfnamefont {A.}~\bibnamefont {Bedoya-Pinto}}, \bibinfo
  {author} {\bibfnamefont {A.}~\bibnamefont {Rizzi}}, \bibinfo {author}
  {\bibfnamefont {C.}~\bibnamefont {Rauch}}, \ and\ \bibinfo {author}
  {\bibfnamefont {F.}~\bibnamefont {Tuomisto}},\ }\href@noop {} {\bibfield
  {journal} {\bibinfo  {journal} {Phys. Rev. B}\ }\textbf {\bibinfo {volume}
  {84}},\ \bibinfo {pages} {081201} (\bibinfo {year} {2011})}\BibitemShut
  {NoStop}%
\bibitem [{\citenamefont {Ramos}\ \emph {et~al.}(2010)\citenamefont {Ramos},
  \citenamefont {Barzola-Quiquia}, \citenamefont {Esquinazi}, \citenamefont
  {Mu\~noz Martin}, \citenamefont {Climent-Font},\ and\ \citenamefont
  {Garcia-Hernandez}}]{PhysRevB.81.214404}%
  \BibitemOpen
  \bibfield  {author} {\bibinfo {author} {\bibfnamefont {M.~A.}\ \bibnamefont
  {Ramos}}, \bibinfo {author} {\bibfnamefont {J.}~\bibnamefont
  {Barzola-Quiquia}}, \bibinfo {author} {\bibfnamefont {P.}~\bibnamefont
  {Esquinazi}}, \bibinfo {author} {\bibfnamefont {A.}~\bibnamefont {Mu\~noz
  Martin}}, \bibinfo {author} {\bibfnamefont {A.}~\bibnamefont {Climent-Font}},
  \ and\ \bibinfo {author} {\bibfnamefont {M.}~\bibnamefont
  {Garcia-Hernandez}},\ }\href {\doibase 10.1103/PhysRevB.81.214404} {\bibfield
   {journal} {\bibinfo  {journal} {Phys. Rev. B}\ }\textbf {\bibinfo {volume}
  {81}},\ \bibinfo {pages} {214404} (\bibinfo {year} {2010})}\BibitemShut
  {NoStop}%
\bibitem [{\citenamefont {Ney}\ \emph {et~al.}(2011)\citenamefont {Ney},
  \citenamefont {Papakonstantinou}, \citenamefont {Kumar}, \citenamefont
  {Shang},\ and\ \citenamefont {Peng}}]{ney2011irradiation}%
  \BibitemOpen
  \bibfield  {author} {\bibinfo {author} {\bibfnamefont {A.}~\bibnamefont
  {Ney}}, \bibinfo {author} {\bibfnamefont {P.}~\bibnamefont
  {Papakonstantinou}}, \bibinfo {author} {\bibfnamefont {A.}~\bibnamefont
  {Kumar}}, \bibinfo {author} {\bibfnamefont {N.-G.}\ \bibnamefont {Shang}}, \
  and\ \bibinfo {author} {\bibfnamefont {N.}~\bibnamefont {Peng}},\ }\href@noop
  {} {\bibfield  {journal} {\bibinfo  {journal} {Appl. Phys. Lett.}\ }\textbf
  {\bibinfo {volume} {99}},\ \bibinfo {pages} {102504} (\bibinfo {year}
  {2011})}\BibitemShut {NoStop}%
\bibitem [{\citenamefont {Nair}\ \emph {et~al.}(2012)\citenamefont {Nair},
  \citenamefont {Sepioni}, \citenamefont {Tsai}, \citenamefont {Lehtinen},
  \citenamefont {Keinonen}, \citenamefont {Krasheninnikov}, \citenamefont
  {Thomson}, \citenamefont {Geim},\ and\ \citenamefont
  {Grigorieva}}]{nair2012spin}%
  \BibitemOpen
  \bibfield  {author} {\bibinfo {author} {\bibfnamefont {R.}~\bibnamefont
  {Nair}}, \bibinfo {author} {\bibfnamefont {M.}~\bibnamefont {Sepioni}},
  \bibinfo {author} {\bibfnamefont {I.-L.}\ \bibnamefont {Tsai}}, \bibinfo
  {author} {\bibfnamefont {O.}~\bibnamefont {Lehtinen}}, \bibinfo {author}
  {\bibfnamefont {J.}~\bibnamefont {Keinonen}}, \bibinfo {author}
  {\bibfnamefont {A.}~\bibnamefont {Krasheninnikov}}, \bibinfo {author}
  {\bibfnamefont {T.}~\bibnamefont {Thomson}}, \bibinfo {author} {\bibfnamefont
  {A.}~\bibnamefont {Geim}}, \ and\ \bibinfo {author} {\bibfnamefont
  {I.}~\bibnamefont {Grigorieva}},\ }\href@noop {} {\bibfield  {journal}
  {\bibinfo  {journal} {Nature Physics}\ }\textbf {\bibinfo {volume} {8}},\
  \bibinfo {pages} {199} (\bibinfo {year} {2012})}\BibitemShut {NoStop}%
\bibitem [{\citenamefont {Sepioni}\ \emph {et~al.}(2010)\citenamefont
  {Sepioni}, \citenamefont {Nair}, \citenamefont {Rablen}, \citenamefont
  {Narayanan}, \citenamefont {Tuna}, \citenamefont {Winpenny}, \citenamefont
  {Geim},\ and\ \citenamefont {Grigorieva}}]{sepioni2010limits}%
  \BibitemOpen
  \bibfield  {author} {\bibinfo {author} {\bibfnamefont {M.}~\bibnamefont
  {Sepioni}}, \bibinfo {author} {\bibfnamefont {R.}~\bibnamefont {Nair}},
  \bibinfo {author} {\bibfnamefont {S.}~\bibnamefont {Rablen}}, \bibinfo
  {author} {\bibfnamefont {J.}~\bibnamefont {Narayanan}}, \bibinfo {author}
  {\bibfnamefont {F.}~\bibnamefont {Tuna}}, \bibinfo {author} {\bibfnamefont
  {R.}~\bibnamefont {Winpenny}}, \bibinfo {author} {\bibfnamefont
  {A.}~\bibnamefont {Geim}}, \ and\ \bibinfo {author} {\bibfnamefont
  {I.}~\bibnamefont {Grigorieva}},\ }\href@noop {} {\bibfield  {journal}
  {\bibinfo  {journal} {Phys. Rev. Lett.}\ }\textbf {\bibinfo {volume} {105}},\
  \bibinfo {pages} {207205} (\bibinfo {year} {2010})}\BibitemShut {NoStop}%
\bibitem [{\citenamefont {He}\ \emph {et~al.}(2012)\citenamefont {He},
  \citenamefont {Yang}, \citenamefont {Xia}, \citenamefont {Regier},
  \citenamefont {Chevrier}, \citenamefont {Zhou},\ and\ \citenamefont
  {Sham}}]{PhysRevB.85.144406}%
  \BibitemOpen
  \bibfield  {author} {\bibinfo {author} {\bibfnamefont {Z.}~\bibnamefont
  {He}}, \bibinfo {author} {\bibfnamefont {X.}~\bibnamefont {Yang}}, \bibinfo
  {author} {\bibfnamefont {H.}~\bibnamefont {Xia}}, \bibinfo {author}
  {\bibfnamefont {T.~Z.}\ \bibnamefont {Regier}}, \bibinfo {author}
  {\bibfnamefont {D.~K.}\ \bibnamefont {Chevrier}}, \bibinfo {author}
  {\bibfnamefont {X.}~\bibnamefont {Zhou}}, \ and\ \bibinfo {author}
  {\bibfnamefont {T.~K.}\ \bibnamefont {Sham}},\ }\href {\doibase
  10.1103/PhysRevB.85.144406} {\bibfield  {journal} {\bibinfo  {journal} {Phys.
  Rev. B}\ }\textbf {\bibinfo {volume} {85}},\ \bibinfo {pages} {144406}
  (\bibinfo {year} {2012})}\BibitemShut {NoStop}%
\bibitem [{\citenamefont {Li}\ \emph {et~al.}(2011)\citenamefont {Li},
  \citenamefont {Prucnal}, \citenamefont {Yao}, \citenamefont {Potzger},
  \citenamefont {Anwand}, \citenamefont {Wagner},\ and\ \citenamefont
  {Zhou}}]{Li201198}%
  \BibitemOpen
  \bibfield  {author} {\bibinfo {author} {\bibfnamefont {L.}~\bibnamefont
  {Li}}, \bibinfo {author} {\bibfnamefont {S.}~\bibnamefont {Prucnal}},
  \bibinfo {author} {\bibfnamefont {S.~D.}\ \bibnamefont {Yao}}, \bibinfo
  {author} {\bibfnamefont {K.}~\bibnamefont {Potzger}}, \bibinfo {author}
  {\bibfnamefont {W.}~\bibnamefont {Anwand}}, \bibinfo {author} {\bibfnamefont
  {A.}~\bibnamefont {Wagner}}, \ and\ \bibinfo {author} {\bibfnamefont
  {S.}~\bibnamefont {Zhou}},\ }\href {\doibase 10.1063/1.3597629} {\bibfield
  {journal} {\bibinfo  {journal} {Appl. Phys. Lett.}\ }\textbf {\bibinfo
  {volume} {98}},\ \bibinfo {eid} {222508} (\bibinfo {year}
  {2011})}\BibitemShut {NoStop}%
\bibitem [{\citenamefont {Zhang}\ \emph {et~al.}(2007)\citenamefont {Zhang},
  \citenamefont {Talapatra}, \citenamefont {Kar}, \citenamefont {Vajtai},
  \citenamefont {Nayak},\ and\ \citenamefont {Ajayan}}]{PhysRevLett.99.107201}%
  \BibitemOpen
  \bibfield  {author} {\bibinfo {author} {\bibfnamefont {Y.}~\bibnamefont
  {Zhang}}, \bibinfo {author} {\bibfnamefont {S.}~\bibnamefont {Talapatra}},
  \bibinfo {author} {\bibfnamefont {S.}~\bibnamefont {Kar}}, \bibinfo {author}
  {\bibfnamefont {R.}~\bibnamefont {Vajtai}}, \bibinfo {author} {\bibfnamefont
  {S.~K.}\ \bibnamefont {Nayak}}, \ and\ \bibinfo {author} {\bibfnamefont
  {P.~M.}\ \bibnamefont {Ajayan}},\ }\href {\doibase
  10.1103/PhysRevLett.99.107201} {\bibfield  {journal} {\bibinfo  {journal}
  {Phys. Rev. Lett.}\ }\textbf {\bibinfo {volume} {99}},\ \bibinfo {pages}
  {107201} (\bibinfo {year} {2007})}\BibitemShut {NoStop}%
\bibitem [{\citenamefont {Ugeda}\ \emph {et~al.}(2010)\citenamefont {Ugeda},
  \citenamefont {Brihuega}, \citenamefont {Guinea},\ and\ \citenamefont
  {Gomez-Rodriguez}}]{PhysRevLett.104.096804}%
  \BibitemOpen
  \bibfield  {author} {\bibinfo {author} {\bibfnamefont {M.~M.}\ \bibnamefont
  {Ugeda}}, \bibinfo {author} {\bibfnamefont {I.}~\bibnamefont {Brihuega}},
  \bibinfo {author} {\bibfnamefont {F.}~\bibnamefont {Guinea}}, \ and\ \bibinfo
  {author} {\bibfnamefont {J.~M.}\ \bibnamefont {Gomez-Rodriguez}},\ }\href
  {\doibase 10.1103/PhysRevLett.104.096804} {\bibfield  {journal} {\bibinfo
  {journal} {Phys. Rev. Lett.}\ }\textbf {\bibinfo {volume} {104}},\ \bibinfo
  {pages} {096804} (\bibinfo {year} {2010})}\BibitemShut {NoStop}%
\bibitem [{\citenamefont {Ohldag}\ \emph {et~al.}(2010)\citenamefont {Ohldag},
  \citenamefont {Esquinazi}, \citenamefont {Arenholz}, \citenamefont {Spemann},
  \citenamefont {Rothermel}, \citenamefont {Setzer},\ and\ \citenamefont
  {Butz}}]{1367-2630-12-12-123012}%
  \BibitemOpen
  \bibfield  {author} {\bibinfo {author} {\bibfnamefont {H.}~\bibnamefont
  {Ohldag}}, \bibinfo {author} {\bibfnamefont {P.}~\bibnamefont {Esquinazi}},
  \bibinfo {author} {\bibfnamefont {E.}~\bibnamefont {Arenholz}}, \bibinfo
  {author} {\bibfnamefont {D.}~\bibnamefont {Spemann}}, \bibinfo {author}
  {\bibfnamefont {M.}~\bibnamefont {Rothermel}}, \bibinfo {author}
  {\bibfnamefont {A.}~\bibnamefont {Setzer}}, \ and\ \bibinfo {author}
  {\bibfnamefont {T.}~\bibnamefont {Butz}},\ }\href@noop {} {\bibfield
  {journal} {\bibinfo  {journal} {New J. Phys.}\ }\textbf {\bibinfo {volume}
  {12}},\ \bibinfo {pages} {123012} (\bibinfo {year} {2010})}\BibitemShut
  {NoStop}%
\bibitem [{\citenamefont {Lehtinen}\ \emph
  {et~al.}(2004{\natexlab{a}})\citenamefont {Lehtinen}, \citenamefont {Foster},
  \citenamefont {Ma}, \citenamefont {Krasheninnikov},\ and\ \citenamefont
  {Nieminen}}]{lehtinen2004irradiation}%
  \BibitemOpen
  \bibfield  {author} {\bibinfo {author} {\bibfnamefont {P.}~\bibnamefont
  {Lehtinen}}, \bibinfo {author} {\bibfnamefont {A.}~\bibnamefont {Foster}},
  \bibinfo {author} {\bibfnamefont {Y.}~\bibnamefont {Ma}}, \bibinfo {author}
  {\bibfnamefont {A.}~\bibnamefont {Krasheninnikov}}, \ and\ \bibinfo {author}
  {\bibfnamefont {R.}~\bibnamefont {Nieminen}},\ }\href@noop {} {\bibfield
  {journal} {\bibinfo  {journal} {Phys. Rev. Lett.}\ }\textbf {\bibinfo
  {volume} {93}},\ \bibinfo {pages} {187202} (\bibinfo {year}
  {2004}{\natexlab{a}})}\BibitemShut {NoStop}%
\bibitem [{\citenamefont {Esquinazi}\ \emph {et~al.}(2010)\citenamefont
  {Esquinazi}, \citenamefont {Barzola-Quiquia}, \citenamefont {Spemann},
  \citenamefont {Rothermel}, \citenamefont {Ohldag}, \citenamefont {Garcia},
  \citenamefont {Setzer},\ and\ \citenamefont {Butz}}]{Esquinazi20101156}%
  \BibitemOpen
  \bibfield  {author} {\bibinfo {author} {\bibfnamefont {P.}~\bibnamefont
  {Esquinazi}}, \bibinfo {author} {\bibfnamefont {J.}~\bibnamefont
  {Barzola-Quiquia}}, \bibinfo {author} {\bibfnamefont {D.}~\bibnamefont
  {Spemann}}, \bibinfo {author} {\bibfnamefont {M.}~\bibnamefont {Rothermel}},
  \bibinfo {author} {\bibfnamefont {H.}~\bibnamefont {Ohldag}}, \bibinfo
  {author} {\bibfnamefont {N.}~\bibnamefont {Garcia}}, \bibinfo {author}
  {\bibfnamefont {A.}~\bibnamefont {Setzer}}, \ and\ \bibinfo {author}
  {\bibfnamefont {T.}~\bibnamefont {Butz}},\ }\href {\doibase
  http://dx.doi.org/10.1016/j.jmmm.2009.06.038} {\bibfield  {journal} {\bibinfo
   {journal} {J. Magn. Magn. Mater.}\ }\textbf {\bibinfo {volume} {322}},\
  \bibinfo {pages} {1156 } (\bibinfo {year} {2010})}\BibitemShut {NoStop}%
\bibitem [{\citenamefont {Sepioni}\ \emph
  {et~al.}(2012{\natexlab{a}})\citenamefont {Sepioni}, \citenamefont {Nair},
  \citenamefont {Tsai}, \citenamefont {Geim},\ and\ \citenamefont
  {Grigorieva}}]{0295-5075-97-4-47001}%
  \BibitemOpen
  \bibfield  {author} {\bibinfo {author} {\bibfnamefont {M.}~\bibnamefont
  {Sepioni}}, \bibinfo {author} {\bibfnamefont {R.~R.}\ \bibnamefont {Nair}},
  \bibinfo {author} {\bibfnamefont {I.-L.}\ \bibnamefont {Tsai}}, \bibinfo
  {author} {\bibfnamefont {A.~K.}\ \bibnamefont {Geim}}, \ and\ \bibinfo
  {author} {\bibfnamefont {I.~V.}\ \bibnamefont {Grigorieva}},\ }\href@noop {}
  {\bibfield  {journal} {\bibinfo  {journal} {EPL (Europhysics Letters)}\
  }\textbf {\bibinfo {volume} {97}},\ \bibinfo {pages} {47001} (\bibinfo {year}
  {2012}{\natexlab{a}})}\BibitemShut {NoStop}%
\bibitem [{\citenamefont {Spemann}\ \emph {et~al.}(2012)\citenamefont
  {Spemann}, \citenamefont {Rothermel}, \citenamefont {Esquinazi},
  \citenamefont {Ramos}, \citenamefont {Kopelevich},\ and\ \citenamefont
  {Ohldag}}]{0295-5075-98-5-57006}%
  \BibitemOpen
  \bibfield  {author} {\bibinfo {author} {\bibfnamefont {D.}~\bibnamefont
  {Spemann}}, \bibinfo {author} {\bibfnamefont {M.}~\bibnamefont {Rothermel}},
  \bibinfo {author} {\bibfnamefont {P.}~\bibnamefont {Esquinazi}}, \bibinfo
  {author} {\bibfnamefont {M.~A.}\ \bibnamefont {Ramos}}, \bibinfo {author}
  {\bibfnamefont {Y.}~\bibnamefont {Kopelevich}}, \ and\ \bibinfo {author}
  {\bibfnamefont {H.}~\bibnamefont {Ohldag}},\ }\href@noop {} {\bibfield
  {journal} {\bibinfo  {journal} {EPL (Europhysics Letters)}\ }\textbf
  {\bibinfo {volume} {98}},\ \bibinfo {pages} {57006} (\bibinfo {year}
  {2012})}\BibitemShut {NoStop}%
\bibitem [{\citenamefont {Sepioni}\ \emph
  {et~al.}(2012{\natexlab{b}})\citenamefont {Sepioni}, \citenamefont {Nair},
  \citenamefont {Tsai}, \citenamefont {Geim}, \citenamefont {Grigorieva} \emph
  {et~al.}}]{sepioni2012reply}%
  \BibitemOpen
  \bibfield  {author} {\bibinfo {author} {\bibfnamefont {M.}~\bibnamefont
  {Sepioni}}, \bibinfo {author} {\bibfnamefont {R.}~\bibnamefont {Nair}},
  \bibinfo {author} {\bibfnamefont {I.}~\bibnamefont {Tsai}}, \bibinfo {author}
  {\bibfnamefont {A.}~\bibnamefont {Geim}}, \bibinfo {author} {\bibfnamefont
  {I.}~\bibnamefont {Grigorieva}},  \emph {et~al.},\ }\href@noop {} {\bibfield
  {journal} {\bibinfo  {journal} {EPL (Europhysics Letters)}\ }\textbf
  {\bibinfo {volume} {98}},\ \bibinfo {pages} {57007} (\bibinfo {year}
  {2012}{\natexlab{b}})}\BibitemShut {NoStop}%
\bibitem [{\citenamefont {Spemann}\ \emph {et~al.}(2013)\citenamefont
  {Spemann}, \citenamefont {Esquinazi}, \citenamefont {Setzer},\ and\
  \citenamefont {B{\"o}hlmann}}]{spemann2013trace}%
  \BibitemOpen
  \bibfield  {author} {\bibinfo {author} {\bibfnamefont {D.}~\bibnamefont
  {Spemann}}, \bibinfo {author} {\bibfnamefont {P.}~\bibnamefont {Esquinazi}},
  \bibinfo {author} {\bibfnamefont {A.}~\bibnamefont {Setzer}}, \ and\ \bibinfo
  {author} {\bibfnamefont {W.}~\bibnamefont {B{\"o}hlmann}},\ }\href@noop {}
  {\bibfield  {journal} {\bibinfo  {journal} {arXiv preprint arXiv:1310.3056}\
  } (\bibinfo {year} {2013})}\BibitemShut {NoStop}%
\bibitem [{\citenamefont {Venkatesan}\ \emph {et~al.}(2013)\citenamefont
  {Venkatesan}, \citenamefont {Dunne}, \citenamefont {Chen}, \citenamefont
  {Zhang},\ and\ \citenamefont {Coey}}]{venkatesan2013structural}%
  \BibitemOpen
  \bibfield  {author} {\bibinfo {author} {\bibfnamefont {M.}~\bibnamefont
  {Venkatesan}}, \bibinfo {author} {\bibfnamefont {P.}~\bibnamefont {Dunne}},
  \bibinfo {author} {\bibfnamefont {Y.}~\bibnamefont {Chen}}, \bibinfo {author}
  {\bibfnamefont {H.}~\bibnamefont {Zhang}}, \ and\ \bibinfo {author}
  {\bibfnamefont {J.}~\bibnamefont {Coey}},\ }\href@noop {} {\bibfield
  {journal} {\bibinfo  {journal} {Carbon}\ }\textbf {\bibinfo {volume} {56}},\
  \bibinfo {pages} {279} (\bibinfo {year} {2013})}\BibitemShut {NoStop}%
\bibitem [{\citenamefont {Sawicki}\ \emph {et~al.}(2011)\citenamefont
  {Sawicki}, \citenamefont {Stefanowicz},\ and\ \citenamefont
  {Ney}}]{sawicki2011sensitive}%
  \BibitemOpen
  \bibfield  {author} {\bibinfo {author} {\bibfnamefont {M.}~\bibnamefont
  {Sawicki}}, \bibinfo {author} {\bibfnamefont {W.}~\bibnamefont
  {Stefanowicz}}, \ and\ \bibinfo {author} {\bibfnamefont {A.}~\bibnamefont
  {Ney}},\ }\href@noop {} {\bibfield  {journal} {\bibinfo  {journal} {Semicond.
  Sci. \& Technol.}\ }\textbf {\bibinfo {volume} {26}},\ \bibinfo {pages}
  {064006} (\bibinfo {year} {2011})}\BibitemShut {NoStop}%
\bibitem [{\citenamefont {Pereira}\ \emph {et~al.}(2011)\citenamefont
  {Pereira}, \citenamefont {Araújo}, \citenamefont {Bael}, \citenamefont
  {Temst},\ and\ \citenamefont {Vantomme}}]{0022-3727-44-21-215001}%
  \BibitemOpen
  \bibfield  {author} {\bibinfo {author} {\bibfnamefont {L.~M.~C.}\
  \bibnamefont {Pereira}}, \bibinfo {author} {\bibfnamefont {J.~P.}\
  \bibnamefont {Araújo}}, \bibinfo {author} {\bibfnamefont {M.~J.~V.}\
  \bibnamefont {Bael}}, \bibinfo {author} {\bibfnamefont {K.}~\bibnamefont
  {Temst}}, \ and\ \bibinfo {author} {\bibfnamefont {A.}~\bibnamefont
  {Vantomme}},\ }\href@noop {} {\bibfield  {journal} {\bibinfo  {journal} {J.
  Phys. D-Appl. Phys.}\ }\textbf {\bibinfo {volume} {44}},\ \bibinfo {pages}
  {215001} (\bibinfo {year} {2011})}\BibitemShut {NoStop}%
\bibitem [{\citenamefont {Lin}\ \emph {et~al.}(2003)\citenamefont {Lin},
  \citenamefont {Alber},\ and\ \citenamefont
  {Henkelmann}}]{lin2003calibration}%
  \BibitemOpen
  \bibfield  {author} {\bibinfo {author} {\bibfnamefont {X.}~\bibnamefont
  {Lin}}, \bibinfo {author} {\bibfnamefont {D.}~\bibnamefont {Alber}}, \ and\
  \bibinfo {author} {\bibfnamefont {R.}~\bibnamefont {Henkelmann}},\
  }\href@noop {} {\bibfield  {journal} {\bibinfo  {journal} {J. Radioanal.
  Nucl. Chem.}\ }\textbf {\bibinfo {volume} {257}},\ \bibinfo {pages} {531}
  (\bibinfo {year} {2003})}\BibitemShut {NoStop}%
\bibitem [{\citenamefont {Wendler}\ \emph {et~al.}(2012)\citenamefont
  {Wendler}, \citenamefont {Bierschenk}, \citenamefont {Felgentr{\"a}ger},
  \citenamefont {Sommerfeld}, \citenamefont {Wesch}, \citenamefont {Alber},
  \citenamefont {Bukalis}, \citenamefont {Prinsloo}, \citenamefont {van~der
  Berg}, \citenamefont {Friedland} \emph {et~al.}}]{wendler2012damage}%
  \BibitemOpen
  \bibfield  {author} {\bibinfo {author} {\bibfnamefont {E.}~\bibnamefont
  {Wendler}}, \bibinfo {author} {\bibfnamefont {T.}~\bibnamefont {Bierschenk}},
  \bibinfo {author} {\bibfnamefont {F.}~\bibnamefont {Felgentr{\"a}ger}},
  \bibinfo {author} {\bibfnamefont {J.}~\bibnamefont {Sommerfeld}}, \bibinfo
  {author} {\bibfnamefont {W.}~\bibnamefont {Wesch}}, \bibinfo {author}
  {\bibfnamefont {D.}~\bibnamefont {Alber}}, \bibinfo {author} {\bibfnamefont
  {G.}~\bibnamefont {Bukalis}}, \bibinfo {author} {\bibfnamefont {L.~C.}\
  \bibnamefont {Prinsloo}}, \bibinfo {author} {\bibfnamefont {N.}~\bibnamefont
  {van~der Berg}}, \bibinfo {author} {\bibfnamefont {E.}~\bibnamefont
  {Friedland}},  \emph {et~al.},\ }\href@noop {} {\bibfield  {journal}
  {\bibinfo  {journal} {Nucl. Instr. Meth. Phys. Res. B}\ }\textbf {\bibinfo
  {volume} {286}},\ \bibinfo {pages} {97} (\bibinfo {year} {2012})}\BibitemShut
  {NoStop}%
\bibitem [{\citenamefont {Kelly}\ \emph {et~al.}(2000)\citenamefont {Kelly},
  \citenamefont {Marsden}, \citenamefont {Hall}, \citenamefont {Martin},
  \citenamefont {Harper},\ and\ \citenamefont
  {Blanchard}}]{kelly2000irradiation}%
  \BibitemOpen
  \bibfield  {author} {\bibinfo {author} {\bibfnamefont {B.}~\bibnamefont
  {Kelly}}, \bibinfo {author} {\bibfnamefont {B.}~\bibnamefont {Marsden}},
  \bibinfo {author} {\bibfnamefont {K.}~\bibnamefont {Hall}}, \bibinfo {author}
  {\bibfnamefont {D.}~\bibnamefont {Martin}}, \bibinfo {author} {\bibfnamefont
  {A.}~\bibnamefont {Harper}}, \ and\ \bibinfo {author} {\bibfnamefont
  {A.}~\bibnamefont {Blanchard}},\ }\href@noop {} {\bibfield  {journal}
  {\bibinfo  {journal} {IAEA Tecdoc}\ }\textbf {\bibinfo {volume} {1154}}
  (\bibinfo {year} {2000})}\BibitemShut {NoStop}%
\bibitem [{\citenamefont {Bode}(1996)}]{bode1996instrumental}%
  \BibitemOpen
  \bibfield  {author} {\bibinfo {author} {\bibfnamefont {P.}~\bibnamefont
  {Bode}},\ }\emph {\bibinfo {title} {Instrumental and organizational aspects
  of a neutron activation analysis laboratory}},\ \href@noop {} {Ph.D.
  thesis},\ \bibinfo  {school} {Delft University of Technology, Delft}
  (\bibinfo {year} {1996})\BibitemShut {NoStop}%
\bibitem [{\citenamefont {Pimenta}\ \emph {et~al.}(2007)\citenamefont
  {Pimenta}, \citenamefont {Dresselhaus}, \citenamefont {Dresselhaus},
  \citenamefont {Cancado}, \citenamefont {Jorio},\ and\ \citenamefont
  {Saito}}]{pimenta2007studying}%
  \BibitemOpen
  \bibfield  {author} {\bibinfo {author} {\bibfnamefont {M.}~\bibnamefont
  {Pimenta}}, \bibinfo {author} {\bibfnamefont {G.}~\bibnamefont
  {Dresselhaus}}, \bibinfo {author} {\bibfnamefont {M.~S.}\ \bibnamefont
  {Dresselhaus}}, \bibinfo {author} {\bibfnamefont {L.}~\bibnamefont
  {Cancado}}, \bibinfo {author} {\bibfnamefont {A.}~\bibnamefont {Jorio}}, \
  and\ \bibinfo {author} {\bibfnamefont {R.}~\bibnamefont {Saito}},\
  }\href@noop {} {\bibfield  {journal} {\bibinfo  {journal} {Phys. Chem. Chem.
  Phys.}\ }\textbf {\bibinfo {volume} {9}},\ \bibinfo {pages} {1276} (\bibinfo
  {year} {2007})}\BibitemShut {NoStop}%
\bibitem [{\citenamefont {Gerber}\ \emph {et~al.}(2010)\citenamefont {Gerber},
  \citenamefont {Krasheninnikov}, \citenamefont {Foster},\ and\ \citenamefont
  {Nieminen}}]{gerber2010first}%
  \BibitemOpen
  \bibfield  {author} {\bibinfo {author} {\bibfnamefont {I.~C.}\ \bibnamefont
  {Gerber}}, \bibinfo {author} {\bibfnamefont {A.~V.}\ \bibnamefont
  {Krasheninnikov}}, \bibinfo {author} {\bibfnamefont {A.~S.}\ \bibnamefont
  {Foster}}, \ and\ \bibinfo {author} {\bibfnamefont {R.~M.}\ \bibnamefont
  {Nieminen}},\ }\href@noop {} {\bibfield  {journal} {\bibinfo  {journal} {New
  J. Phys.}\ }\textbf {\bibinfo {volume} {12}},\ \bibinfo {pages} {113021}
  (\bibinfo {year} {2010})}\BibitemShut {NoStop}%
\bibitem [{\citenamefont {Arrott}(1957)}]{arrott1957criterion}%
  \BibitemOpen
  \bibfield  {author} {\bibinfo {author} {\bibfnamefont {A.}~\bibnamefont
  {Arrott}},\ }\href@noop {} {\bibfield  {journal} {\bibinfo  {journal} {Phys.
  Rev.}\ }\textbf {\bibinfo {volume} {108}},\ \bibinfo {pages} {1394} (\bibinfo
  {year} {1957})}\BibitemShut {NoStop}%
\bibitem [{\citenamefont {Wohlfarth}(1968)}]{wohlfarth1968very}%
  \BibitemOpen
  \bibfield  {author} {\bibinfo {author} {\bibfnamefont {E.}~\bibnamefont
  {Wohlfarth}},\ }\href@noop {} {\bibfield  {journal} {\bibinfo  {journal} {J.
  Appl. Phys.}\ }\textbf {\bibinfo {volume} {39}},\ \bibinfo {pages} {1061}
  (\bibinfo {year} {1968})}\BibitemShut {NoStop}%
\bibitem [{\citenamefont {He}\ \emph {et~al.}(2011{\natexlab{b}})\citenamefont
  {He}, \citenamefont {Xia}, \citenamefont {Zhou}, \citenamefont {Yang},
  \citenamefont {Song},\ and\ \citenamefont {Wang}}]{he2011raman}%
  \BibitemOpen
  \bibfield  {author} {\bibinfo {author} {\bibfnamefont {Z.}~\bibnamefont
  {He}}, \bibinfo {author} {\bibfnamefont {H.}~\bibnamefont {Xia}}, \bibinfo
  {author} {\bibfnamefont {X.}~\bibnamefont {Zhou}}, \bibinfo {author}
  {\bibfnamefont {X.}~\bibnamefont {Yang}}, \bibinfo {author} {\bibfnamefont
  {Y.}~\bibnamefont {Song}}, \ and\ \bibinfo {author} {\bibfnamefont
  {T.}~\bibnamefont {Wang}},\ }\href@noop {} {\bibfield  {journal} {\bibinfo
  {journal} {J. Phys. D: Appl. Phys.}\ }\textbf {\bibinfo {volume} {44}},\
  \bibinfo {pages} {085001} (\bibinfo {year} {2011}{\natexlab{b}})}\BibitemShut
  {NoStop}%
\bibitem [{\citenamefont {Elman}\ \emph {et~al.}(1982)\citenamefont {Elman},
  \citenamefont {Shayegan}, \citenamefont {Dresselhaus}, \citenamefont
  {Mazurek},\ and\ \citenamefont {Dresselhaus}}]{elman1982structural}%
  \BibitemOpen
  \bibfield  {author} {\bibinfo {author} {\bibfnamefont {B.}~\bibnamefont
  {Elman}}, \bibinfo {author} {\bibfnamefont {M.}~\bibnamefont {Shayegan}},
  \bibinfo {author} {\bibfnamefont {M.}~\bibnamefont {Dresselhaus}}, \bibinfo
  {author} {\bibfnamefont {H.}~\bibnamefont {Mazurek}}, \ and\ \bibinfo
  {author} {\bibfnamefont {G.}~\bibnamefont {Dresselhaus}},\ }\href@noop {}
  {\bibfield  {journal} {\bibinfo  {journal} {Phys. Rev. B}\ }\textbf {\bibinfo
  {volume} {25}},\ \bibinfo {pages} {4142} (\bibinfo {year}
  {1982})}\BibitemShut {NoStop}%
\bibitem [{\citenamefont {Xing}\ \emph {et~al.}(2013)\citenamefont {Xing},
  \citenamefont {Li}, \citenamefont {Hou}, \citenamefont {Hu}, \citenamefont
  {Zhou}, \citenamefont {Peter}, \citenamefont {Petravic},\ and\ \citenamefont
  {Chen}}]{xing2013disorder}%
  \BibitemOpen
  \bibfield  {author} {\bibinfo {author} {\bibfnamefont {T.}~\bibnamefont
  {Xing}}, \bibinfo {author} {\bibfnamefont {L.~H.}\ \bibnamefont {Li}},
  \bibinfo {author} {\bibfnamefont {L.}~\bibnamefont {Hou}}, \bibinfo {author}
  {\bibfnamefont {X.}~\bibnamefont {Hu}}, \bibinfo {author} {\bibfnamefont
  {S.}~\bibnamefont {Zhou}}, \bibinfo {author} {\bibfnamefont {R.}~\bibnamefont
  {Peter}}, \bibinfo {author} {\bibfnamefont {M.}~\bibnamefont {Petravic}}, \
  and\ \bibinfo {author} {\bibfnamefont {Y.}~\bibnamefont {Chen}},\ }\href@noop
  {} {\bibfield  {journal} {\bibinfo  {journal} {Carbon}\ }\textbf {\bibinfo
  {volume} {57}},\ \bibinfo {pages} {515} (\bibinfo {year} {2013})}\BibitemShut
  {NoStop}%
\bibitem [{\citenamefont {Telling}\ and\ \citenamefont
  {Heggie}(2007)}]{telling2007radiation}%
  \BibitemOpen
  \bibfield  {author} {\bibinfo {author} {\bibfnamefont {R.}~\bibnamefont
  {Telling}}\ and\ \bibinfo {author} {\bibfnamefont {M.}~\bibnamefont
  {Heggie}},\ }\href@noop {} {\bibfield  {journal} {\bibinfo  {journal}
  {Philosophical Magazine}\ }\textbf {\bibinfo {volume} {87}},\ \bibinfo
  {pages} {4797} (\bibinfo {year} {2007})}\BibitemShut {NoStop}%
\bibitem [{\citenamefont {Makarova}\ \emph {et~al.}(2008)\citenamefont
  {Makarova}, \citenamefont {Ricc{\`o}}, \citenamefont {Pontiroli},
  \citenamefont {Mazzani}, \citenamefont {Belli},\ and\ \citenamefont
  {Goffredi}}]{makarova2008ageing}%
  \BibitemOpen
  \bibfield  {author} {\bibinfo {author} {\bibfnamefont {T.}~\bibnamefont
  {Makarova}}, \bibinfo {author} {\bibfnamefont {M.}~\bibnamefont {Ricc{\`o}}},
  \bibinfo {author} {\bibfnamefont {D.}~\bibnamefont {Pontiroli}}, \bibinfo
  {author} {\bibfnamefont {M.}~\bibnamefont {Mazzani}}, \bibinfo {author}
  {\bibfnamefont {M.}~\bibnamefont {Belli}}, \ and\ \bibinfo {author}
  {\bibfnamefont {A.}~\bibnamefont {Goffredi}},\ }\href@noop {} {\bibfield
  {journal} {\bibinfo  {journal} {physica status solidi (b)}\ }\textbf
  {\bibinfo {volume} {245}},\ \bibinfo {pages} {2082} (\bibinfo {year}
  {2008})}\BibitemShut {NoStop}%
\bibitem [{\citenamefont {Jawhari}\ \emph {et~al.}(1995)\citenamefont
  {Jawhari}, \citenamefont {Roid},\ and\ \citenamefont
  {Casado}}]{jawhari1995raman}%
  \BibitemOpen
  \bibfield  {author} {\bibinfo {author} {\bibfnamefont {T.}~\bibnamefont
  {Jawhari}}, \bibinfo {author} {\bibfnamefont {A.}~\bibnamefont {Roid}}, \
  and\ \bibinfo {author} {\bibfnamefont {J.}~\bibnamefont {Casado}},\
  }\href@noop {} {\bibfield  {journal} {\bibinfo  {journal} {Carbon}\ }\textbf
  {\bibinfo {volume} {33}},\ \bibinfo {pages} {1561} (\bibinfo {year}
  {1995})}\BibitemShut {NoStop}%
\bibitem [{\citenamefont {Ohldag}\ \emph {et~al.}(2007)\citenamefont {Ohldag},
  \citenamefont {Tyliszczak}, \citenamefont {H\"ohne}, \citenamefont {Spemann},
  \citenamefont {Esquinazi}, \citenamefont {Ungureanu},\ and\ \citenamefont
  {Butz}}]{PhysRevLett.98.187204}%
  \BibitemOpen
  \bibfield  {author} {\bibinfo {author} {\bibfnamefont {H.}~\bibnamefont
  {Ohldag}}, \bibinfo {author} {\bibfnamefont {T.}~\bibnamefont {Tyliszczak}},
  \bibinfo {author} {\bibfnamefont {R.}~\bibnamefont {H\"ohne}}, \bibinfo
  {author} {\bibfnamefont {D.}~\bibnamefont {Spemann}}, \bibinfo {author}
  {\bibfnamefont {P.}~\bibnamefont {Esquinazi}}, \bibinfo {author}
  {\bibfnamefont {M.}~\bibnamefont {Ungureanu}}, \ and\ \bibinfo {author}
  {\bibfnamefont {T.}~\bibnamefont {Butz}},\ }\href {\doibase
  10.1103/PhysRevLett.98.187204} {\bibfield  {journal} {\bibinfo  {journal}
  {Phys. Rev. Lett.}\ }\textbf {\bibinfo {volume} {98}},\ \bibinfo {pages}
  {187204} (\bibinfo {year} {2007})}\BibitemShut {NoStop}%
\bibitem [{\citenamefont {Lehtinen}\ \emph
  {et~al.}(2004{\natexlab{b}})\citenamefont {Lehtinen}, \citenamefont {Foster},
  \citenamefont {Ma}, \citenamefont {Krasheninnikov},\ and\ \citenamefont
  {Nieminen}}]{PhysRevLett.93.187202}%
  \BibitemOpen
  \bibfield  {author} {\bibinfo {author} {\bibfnamefont {P.~O.}\ \bibnamefont
  {Lehtinen}}, \bibinfo {author} {\bibfnamefont {A.~S.}\ \bibnamefont
  {Foster}}, \bibinfo {author} {\bibfnamefont {Y.}~\bibnamefont {Ma}}, \bibinfo
  {author} {\bibfnamefont {A.~V.}\ \bibnamefont {Krasheninnikov}}, \ and\
  \bibinfo {author} {\bibfnamefont {R.~M.}\ \bibnamefont {Nieminen}},\ }\href
  {\doibase 10.1103/PhysRevLett.93.187202} {\bibfield  {journal} {\bibinfo
  {journal} {Phys. Rev. Lett.}\ }\textbf {\bibinfo {volume} {93}},\ \bibinfo
  {pages} {187202} (\bibinfo {year} {2004}{\natexlab{b}})}\BibitemShut
  {NoStop}%
\bibitem [{\citenamefont {Yazyev}(2008)}]{yazyev2008magnetism}%
  \BibitemOpen
  \bibfield  {author} {\bibinfo {author} {\bibfnamefont {O.~V.}\ \bibnamefont
  {Yazyev}},\ }\href@noop {} {\bibfield  {journal} {\bibinfo  {journal} {Phys.
  Rev. Lett.}\ }\textbf {\bibinfo {volume} {101}},\ \bibinfo {pages} {037203}
  (\bibinfo {year} {2008})}\BibitemShut {NoStop}%
\bibitem [{\citenamefont {Telling}\ \emph {et~al.}(2003)\citenamefont
  {Telling}, \citenamefont {Ewels}, \citenamefont {El-Barbary},\ and\
  \citenamefont {Heggie}}]{Telling2003}%
  \BibitemOpen
  \bibfield  {author} {\bibinfo {author} {\bibfnamefont {R.}~\bibnamefont
  {Telling}}, \bibinfo {author} {\bibfnamefont {C.}~\bibnamefont {Ewels}},
  \bibinfo {author} {\bibfnamefont {A.}~\bibnamefont {El-Barbary}}, \ and\
  \bibinfo {author} {\bibfnamefont {M.}~\bibnamefont {Heggie}},\ }\href
  {\doibase 10.1038/nmat876} {\bibfield  {journal} {\bibinfo  {journal} {Nat.
  Mater.}\ }\textbf {\bibinfo {volume} {2}},\ \bibinfo {pages} {333} (\bibinfo
  {year} {2003})}\BibitemShut {NoStop}%
\bibitem [{\citenamefont {Machado-Charry}\ \emph {et~al.}(2012)\citenamefont
  {Machado-Charry}, \citenamefont {Boulanger}, \citenamefont {Genovese},
  \citenamefont {Mousseau},\ and\ \citenamefont {Pochet}}]{Machado2012}%
  \BibitemOpen
  \bibfield  {author} {\bibinfo {author} {\bibfnamefont {E.}~\bibnamefont
  {Machado-Charry}}, \bibinfo {author} {\bibfnamefont {P.}~\bibnamefont
  {Boulanger}}, \bibinfo {author} {\bibfnamefont {L.}~\bibnamefont {Genovese}},
  \bibinfo {author} {\bibfnamefont {N.}~\bibnamefont {Mousseau}}, \ and\
  \bibinfo {author} {\bibfnamefont {P.}~\bibnamefont {Pochet}},\ }\href
  {\doibase http://dx.doi.org/10.1063/1.4754143} {\bibfield  {journal}
  {\bibinfo  {journal} {Appl. Phys. Lett.}\ }\textbf {\bibinfo {volume}
  {101}},\ \bibinfo {eid} {132405} (\bibinfo {year} {2012})}\BibitemShut
  {NoStop}%
\bibitem [{\citenamefont {Krishnan}\ \emph {et~al.}(2013)\citenamefont
  {Krishnan}, \citenamefont {Brenet}, \citenamefont {Machado-Charry},
  \citenamefont {Caliste}, \citenamefont {Genovese}, \citenamefont {Deutsch},\
  and\ \citenamefont {Pochet}}]{Krishnan2013}%
  \BibitemOpen
  \bibfield  {author} {\bibinfo {author} {\bibfnamefont {S.}~\bibnamefont
  {Krishnan}}, \bibinfo {author} {\bibfnamefont {G.}~\bibnamefont {Brenet}},
  \bibinfo {author} {\bibfnamefont {E.}~\bibnamefont {Machado-Charry}},
  \bibinfo {author} {\bibfnamefont {D.}~\bibnamefont {Caliste}}, \bibinfo
  {author} {\bibfnamefont {L.}~\bibnamefont {Genovese}}, \bibinfo {author}
  {\bibfnamefont {T.}~\bibnamefont {Deutsch}}, \ and\ \bibinfo {author}
  {\bibfnamefont {P.}~\bibnamefont {Pochet}},\ }\href
  {http://scitation.aip.org/content/aip/journal/apl/103/25/10.1063/1.4850877}
  {\bibfield  {journal} {\bibinfo  {journal} {Appl. Phys. Lett.}\ }\textbf
  {\bibinfo {volume} {103}},\ \bibinfo {eid} {251904} (\bibinfo {year}
  {2013})}\BibitemShut {NoStop}%
\bibitem [{\citenamefont {Genovese}\ \emph {et~al.}(2008)\citenamefont
  {Genovese}, \citenamefont {Neelov}, \citenamefont {Goedecker}, \citenamefont
  {Deutsch}, \citenamefont {Ghasemi}, \citenamefont {Willand}, \citenamefont
  {Caliste}, \citenamefont {Zilberberg}, \citenamefont {Rayson}, \citenamefont
  {Bergman},\ and\ \citenamefont {Schneider}}]{Genovese2008}%
  \BibitemOpen
  \bibfield  {author} {\bibinfo {author} {\bibfnamefont {L.}~\bibnamefont
  {Genovese}}, \bibinfo {author} {\bibfnamefont {A.}~\bibnamefont {Neelov}},
  \bibinfo {author} {\bibfnamefont {S.}~\bibnamefont {Goedecker}}, \bibinfo
  {author} {\bibfnamefont {T.}~\bibnamefont {Deutsch}}, \bibinfo {author}
  {\bibfnamefont {S.~A.}\ \bibnamefont {Ghasemi}}, \bibinfo {author}
  {\bibfnamefont {A.}~\bibnamefont {Willand}}, \bibinfo {author} {\bibfnamefont
  {D.}~\bibnamefont {Caliste}}, \bibinfo {author} {\bibfnamefont
  {O.}~\bibnamefont {Zilberberg}}, \bibinfo {author} {\bibfnamefont
  {M.}~\bibnamefont {Rayson}}, \bibinfo {author} {\bibfnamefont
  {A.}~\bibnamefont {Bergman}}, \ and\ \bibinfo {author} {\bibfnamefont
  {R.}~\bibnamefont {Schneider}},\ }\href@noop {} {\bibfield  {journal}
  {\bibinfo  {journal} {J. Chem. Phys.}\ }\textbf {\bibinfo {volume} {129}}
  (\bibinfo {year} {2008})}\BibitemShut {NoStop}%
\bibitem [{\citenamefont {Genovese}\ \emph {et~al.}(2007)\citenamefont
  {Genovese}, \citenamefont {Deutsch},\ and\ \citenamefont
  {Goedecker}}]{Genovese2007}%
  \BibitemOpen
  \bibfield  {author} {\bibinfo {author} {\bibfnamefont {L.}~\bibnamefont
  {Genovese}}, \bibinfo {author} {\bibfnamefont {T.}~\bibnamefont {Deutsch}}, \
  and\ \bibinfo {author} {\bibfnamefont {S.}~\bibnamefont {Goedecker}},\
  }\href@noop {} {\bibfield  {journal} {\bibinfo  {journal} {J. Chem. Phys.}\
  }\textbf {\bibinfo {volume} {127}} (\bibinfo {year} {2007})}\BibitemShut
  {NoStop}%
\bibitem [{\citenamefont {Yazyev}\ and\ \citenamefont
  {Helm}(2007)}]{PhysRevB75125408}%
  \BibitemOpen
  \bibfield  {author} {\bibinfo {author} {\bibfnamefont {O.~V.}\ \bibnamefont
  {Yazyev}}\ and\ \bibinfo {author} {\bibfnamefont {L.}~\bibnamefont {Helm}},\
  }\href {\doibase 10.1103/PhysRevB.75.125408} {\bibfield  {journal} {\bibinfo
  {journal} {Phys. Rev. B}\ }\textbf {\bibinfo {volume} {75}},\ \bibinfo
  {pages} {125408} (\bibinfo {year} {2007})}\BibitemShut {NoStop}%
\bibitem [{\citenamefont {Barzola-Quiquia}\ \emph {et~al.}(2007)\citenamefont
  {Barzola-Quiquia}, \citenamefont {Esquinazi}, \citenamefont {Rothermel},
  \citenamefont {Spemann}, \citenamefont {Butz},\ and\ \citenamefont
  {Garc\'{i}a}}]{PhysRevB.76.161403}%
  \BibitemOpen
  \bibfield  {author} {\bibinfo {author} {\bibfnamefont {J.}~\bibnamefont
  {Barzola-Quiquia}}, \bibinfo {author} {\bibfnamefont {P.}~\bibnamefont
  {Esquinazi}}, \bibinfo {author} {\bibfnamefont {M.}~\bibnamefont
  {Rothermel}}, \bibinfo {author} {\bibfnamefont {D.}~\bibnamefont {Spemann}},
  \bibinfo {author} {\bibfnamefont {T.}~\bibnamefont {Butz}}, \ and\ \bibinfo
  {author} {\bibfnamefont {N.}~\bibnamefont {Garc\'{i}a}},\ }\href {\doibase
  10.1103/PhysRevB.76.161403} {\bibfield  {journal} {\bibinfo  {journal} {Phys.
  Rev. B}\ }\textbf {\bibinfo {volume} {76}},\ \bibinfo {pages} {161403}
  (\bibinfo {year} {2007})}\BibitemShut {NoStop}%
\end{thebibliography}

%

\end{document}